\documentclass[12pt]{article}

\usepackage{amsmath,amssymb}

\newcommand{\ident}{{\rm 1}\!\!\,{\rm l}}

\newcommand{\beq}{\begin{eqnarray}}
\newcommand{\eeq}{\end{eqnarray}}

\begin{document}

\thispagestyle{empty}

\begin{center}


\begin{flushright}
BCCUNY-HEP/06-02
\end{flushright}

\vspace{15pt}
{\large \bf INTEGRABLE MODELS AND CONFINEMENT IN
$(2+1)$-DIMENSIONAL WEAKLY-COUPLED YANG-MILLS THEORY  }

\vspace{20pt}

{\bf Peter Orland}$^{\rm a.b.c.}$\!\footnote{giantswing@gursey.baruch.cuny.edu}

\vspace{8pt}

\begin{flushleft}
a. Physics Program, The Graduate School and University Center,
The City University of New York, 365 Fifth Avenue,
New York, NY 10016, U.S.A.
\end{flushleft}

\begin{flushleft}
b. Department of Natural Sciences, Baruch College, The 
City University of New York, 17 Lexington Avenue, New 
York, NY 10010, U.S.A. 
\end{flushleft}

\begin{flushleft}
c. The Niels Bohr Institute, Blegamsvej 17, DK-2100, Copenhagen {\O}, Denmark
 \end{flushleft}

\vspace{40pt}

{\bf Abstract}
\end{center}

\noindent 
We generalize the $(2+1)$-dimensional Yang-Mills theory 
to an anisotropic form with two gauge
coupling constants $e$ and $e^{\prime}$. In axial gauge, a regularized version of
the Hamiltonian of this gauge theory 
is $H_{0}+{e^{\prime}}^{2}H_{1}$, where $H_{0}$ is the Hamiltonian of a set of 
$(1+1)$-dimensional principal chiral sigma nonlinear
models and $H_{1}$ 
couples charge densities of these sigma models. We treat
$H_{1}$ as the interaction Hamiltonian. For gauge group SU($2$), we use form
factors of the currents of the principal chiral sigma models to compute the string tension 
for small $e^{\prime}$,  after reviewing exact S-matrix and 
form-factor methods. In the anisotropic regime, the dependence of 
the string tension on the coupling constants 
is not in accord with generally-accepted dimensional arguments.

\hfill

\newpage

\setcounter{footnote}{0}

\setcounter{page}{1}

\section{Introduction}
\setcounter{equation}{0}
\renewcommand{\theequation}{1.\arabic{equation}}

In this paper we calculate the string tension of pure $(2+1)$-dimensional Yang-Mills theory at
weak coupling. This is done in an anisotropic version of the theory, with two different, but
small, coupling constants. The result clearly establishes that confinement is
not restricted to the strong-coupling region. 

The method used is a development of the proposal of reference \cite{PhysRevD71}, which
in turn was inspired by Mandelstam's arguments for confinement in 
QCD \cite{mandelstam}. The starting
point is a regularized Hamiltonian in an axial gauge, $A_{1}=0$, where $A_{j}, j=1,2$ is
the Lie-algebra valued gauge field (the component $A_{0}$ is also set to zero). The 
Hamiltonian
may be written as a sum of two terms, namely
\beq
H_{0}=\int d^{2}x\left( \frac{e^{2}}{2} {\rm Tr}\,{\mathcal E}_{2}^{2} 
+\frac{1}{2e^{2}}{\rm Tr}\,{\mathcal B}^{2} \right) \;,\nonumber
\eeq
and
\beq
H_{1}=\frac{e^{2}}{2}\int d^{2}x  {\rm Tr}\,{\mathcal E}_{1}^{2}\;, \nonumber
\eeq 
where ${\mathcal E}_{j}$, $j=1,2$ are the components of the electric field and $\mathcal B$ is
the magnetic field
${\mathcal B}={\rm i}[\partial_{1}-{\rm  i}A_{1},\partial_{2}-{\rm i}A_{2}]=\partial_{1}A_{2}$  Though we
may write ${\mathcal E}_{2} =-{\rm i}\delta/\delta A_{2}$, in the Schr\"{o}dinger representation,
the formula for ${\mathcal E}_{1}$ is a nonlocal expression, obtained from Gauss's law 
\cite{mandelstam}
\begin{eqnarray}
{\mathcal E}_{1}(x) 
&\!=\!&
-\int^{x^{1}} \!dy^{1} \;
[\partial_{2}-iA_{2}(y^{1},x^{2}), {\mathcal E}_{2}(y^{1},x^{2})] \nonumber \\
&\!=\!&-\int^{x^{1}} \! dy^{1}\;
{\mathcal D}_{2}(y^{1},x^{\perp}) \cdot
{\mathcal E}_{2}(y^{1}, x^{2})
\;, \label{continuum-electric}
\end{eqnarray}
where ${\mathcal D}_{2}$ is the adjoint covariant derivative in the two-direction. The local
form of Gauss's law is explicitly satisfied with (\ref{continuum-electric}), provided a residual
gauge invariance 
\beq
\int dx^{1} {\mathcal D}_{2}{\mathcal E}_{2} \;\Psi=0 \;,\label{continuum-physical}
\eeq
is satisfied by physical states $\Psi$ (this condition must be modified slightly if
quarks are present).

If the theory is regularized on a lattice, $H_{0}$ is the Hamiltonian for a set of
decoupled $\rm{SU}(N) \!\times\! {\rm SU}(N)$ principal chiral nonlinear
sigma models. These sigma
models are coupled together by the interaction Hamiltonian $H_{1}$. We shall treat the
coefficient of $H_{1}$ as small, meaning that we consider an anisotropic 
modification, 
\beq
H_{1}=\frac{(e^{\prime})^{2}}{2}\int d^{2}x  {\rm Tr}\,{\mathcal E}_{1}^{2}\;. \nonumber
\eeq

By assuming $e^{\prime}\ll e$, simple arguments were made that the theory is
in a confining phase \cite{PhysRevD71}. It
was also suggested that Rayleigh-Schr\"{o}dinger perturbation theory in $e^{\prime}$ was 
infrared
finite - but it is now clear to the author that this is not the case. A better understanding
is needed to obtain quantitative results.

In this paper, we shall work with gauge group
SU($2$) and exploit exact
knowledge of the form factors of the $(1+1)$-dimensional
${\rm O}(4)\simeq {\rm SU}(2)\!\times\! {\rm SU}(2)$
nonlinear sigma model \cite{KarowskiWeisz}. The expressions for the
form factors are valid for
small $e$, as we shall show, so 
both gauge couplings must be small. After reintroducing the component
$A_{0}$ of the gauge field, we integrate out $A_{2}$ using these form factors. We determine 
the resulting
effective action of $A_{0}$ to leading order. This effective action is infrared finite.

We shall find a result for the string
tension in the $x^{1}$-direction which is different from that in reference \cite{PhysRevD71}
and will discuss how the difference comes about. We do not calculate the string
tension in the $x^{2}$-direction here, which is still under investigation.

In the course of these investigations, we discovered a paper by Griffin which suggests
a program very similar to ours \cite{Griffin}. Evidently, his ideas were taken no further. Griffin's 
proposal was 
for the light-cone gauge, instead
of the axial gauge, but the two are very similar in certain respects. 

We should say a few words about other analytic
approaches to Yang-Mills theories in 
$(2+1)$-dimensions. Probably the best known paper is that of 
Feynman \cite{feynman}. Though
some of the mathematical 
details of Feynman's paper are not correct, evidence 
for some of his conjectures
were found in reference \cite{orl-sem}, which utilized some methods introduced in reference 
\cite{orbit-space}. In particular, the low-magnetic-energy region of configuration space
was argued to be bounded in $(2+1)$ dimensions (it is unbounded in $(3+1)$ dimensions
\cite{orbit-space}), which is indicative of a mass gap. 

An approach, which at first appears
rather different, was developed by Karabali and Nair \cite{kar-nair} in which 
new coordinates for
configuration space were devised. A re-derivation
of one form of the Hamiltonian discussed in reference \cite{kar-nair} was done in
\cite{orl-gauge-inv} using a kind of non-Abelian exterior differentiation; in this way 
a connection
was made with the ideas of references \cite{orbit-space} and \cite{orl-sem}. 

The major achievement of 
reference \cite{kar-nair}
was the finding of a mass gap at strong coupling and a {\em resummation} of the 
strong-coupling expansion. This resummation was used
to obtain a string tension which is independent of the ultraviolet cut-off. We should
mention, however, that very similar results can be obtained by analytic lattice methods. The 
strong-coupling spectrum in the lattice Hamiltonian formalism has the same dependence on
the continuum coupling as in Karabali and Nair's formalism. A
different resummation method on the lattice, due to Greensite \cite{Greensite}
also yields the strong-coupling
vacuum wave functional. The form of Greensite's  vacuum wave functional 
in the continuum limit in $2+1$ dimensions is
\beq
\Psi_{0}=\exp -\frac{1}{4(N-1/N)e^{4}}\int d^{2} x {\rm Tr} [F_{ij}(x)]^{2}\;. \nonumber
\eeq
The wave functional that Karabali and Nair obtained has the form, in the infrared limit,
\beq
\Psi_{0}=\exp -\frac{\pi}{C_{A}e^{4}}\int d^{2} x {\rm Tr} [F_{ij}(x)]^{2}\;, \nonumber
\eeq
where $C_{A}$ is the quadratic Casimir for the adjoint representation. Both vacua yield  
a string tension proportional to $e^{4}$. It 
does not seem to have been noticed before that the essential features of Karabali and Nair's
strong-coupling expansion and those of the lattice strong-coupling expansion worked
out by Greensite are the same. 

To obtain a genuine proof of confinement though strong-coupling methods, it must
be shown that the strong-coupling expansion has a finite radius of convergence in
the $1/e$ or an infinte radius of convergence in the dimensionless coupling
$1/g_{0}$. This has not been accomplished yet. 

Recently a new approach
has appeared \cite{leigh-min-yel}, which uses a scheme to improve a Gaussian {\em Ansatz} 
for the vacuum and excited-state wave functionals
in the variables of reference \cite{kar-nair}.

What distinguishes the approach of  \cite{PhysRevD71} and this paper 
from other work is that does not exploit
strong-coupling approximations or any {\em Ansatz} for wave functionals. We do make 
assumptions, 
which are quite different from the assumption of reference \cite{leigh-min-yel}. We assume that
an anisotropic weak-coupling expansion makes sense and
that the result for the ${\rm SU}(2)\! \times \! {\rm SU}(2)$ principal-chiral-sigma-model 
exact form factor \cite{KarowskiWeisz} (which has been checked in the $1/N$ expansion for
the O($4$) formulation) is correct. We think that our approach leaves no
doubt that confinement occurs at weak coupling, granting that
the anisotropy is something we would like to get 
beyond. Or perhaps not - the anisotropic theory is asymptotically 
free and not finite like the
isotropic theory. In this respect it is more like
real QCD in 3+1 dimensions.

By treating the coefficient
of ${\rm Tr} {\mathcal E}_{1}^{2}$ (instead of ${\rm Tr}{\mathcal B}^{2}$) as small, our method is
inherently a weak-coupling approach. We think much insight can be
gained by working systematically 
at small coupling. Furthermore, to solve the much 
harder problem of
QCD in $(3+1)$ dimensions, a weak-coupling understanding is 
essential.

There is a well-accepted argument concerning how the mass gap $M$ and string 
tension $\sigma$ depend
on the coupling constant (see for example references \cite{kar-nair} and \cite{teper}). We find that such an argument fails in the anisotropic
weak-coupling regime. The argument goes as follows: Yang-Mills theory is 
a perturbatively-ultraviolet-finite field theory (there are severe infrared divergences, but
let us ignore these). Thus, after being suitably regularized, the coupling has
a nonzero finite value, as the ultraviolet regulator is removed. Since the coupling squared $e^{2}$,
has units of $cm^{-1}$, the mass gap must behave as $M\sim e^{2}$ and the
string tension must behave as $\sigma \sim e^{4}$. In our anisotropic case, the same dimensional
reasoning implies 
\beq
M\sim \sum_{p}C_{p}\,e^{2-p}(e^{\prime})^{p}\;,\;\;
\sigma \sim \sum_{P}K_{P}\, e^{4-P}(e^{\prime})^{P}\;, \label{naive}
\eeq
for some set of numbers $p$ and $P$ and dimensionless constants
$C_{p}$ and $K_{P}$.  Our final answer for the
string tension is quite different from (\ref{naive}). We
show in Section 6 that for two quarks separated in the $x^{1}$-direction,
\beq
\sigma \sim \frac{e^{2}}{a} \exp-\frac{4\pi}{e^{2}a} \;, \nonumber
\eeq
where $a$ is a short-distance cut-off. We believe that
for $e=e^{\prime}$ the dimensional 
argument does yield the right answer, but that there is a crossover
phenomenon between (\ref{naive}) to the behavior we find in the anisotropic regime.

Another application of exact form factors to the
$(2+1)$-dimensional SU($2$) gauge theory has just appeared
\cite{CaselleGrinzaMagnoli}. In this work, the 
form factors of the two-dimensional Ising model are used to
find the profile of the electric string, near the high-temperature deconfining 
transition, assuming the Svetitsky-Yaffe conjecture.

An interesting question is the value of k-string tensions for gauge group SU($N$)
(see reference \cite{GreensiteReview} for detailed review of this 
matter). This is not discussed in this paper, since the gauge group is
SU(2). Thus the value of k is always one.  Another question is whether adjoint sources
are confined. Both of these issues are addressed in a new paper \cite{newpaper}. The sine 
law is clearly seen for the vertical string tensions. The situation is less clear for horizontal string
tensions; at zeroth order in $g_{0}^{\prime}$
there is a Casimir law, but there are corrections. We are unable to 
calculate these corrections, because 
we do not know the form factors for ${\rm SU}(N)\! \times \! {\rm SU}(N)$ principal chiral sigma 
models. Adjoint sources are shown not to be confined.

In the next section, we discuss the regularized Hamiltonian. In Section 3, we go to
the axial gauge and
split this Hamiltonian into the Hamiltonians of $(1+1)$-dimensional
${\rm O}(4)\simeq {\rm SU}(2)\!\times\! {\rm SU}(2)$
nonlinear sigma models $H_{0}$ and
a nonlocal term $H_{1}$. We discuss how to find the effective
action of the temporal gauge field in terms of correlators of the nonlinear 
sigma model in Section 4.  In 
Section 5, we determine the leading-order effective action using
the exact form factors of the O($N$) nonlinear sigma model in $(1+1)$-dimensions. The
static potential is then found between two quarks separated in the $x^{1}$-direction in Section 6. The
physical picture of confinement of glueball excitations 
and quarks separated in the $x^{2}$-direction is presented
in Section 7. We discuss
some future endeavors in Section 8. We give a review for non-experts
on the exact S-matrix \cite{Zamolodchikov}
and form factors \cite{KarowskiWeisz} of the 
$(1+1)$-dimensional
${\rm O}(N)$ nonlinear sigma model in the Appendix.

\section{The regularized Hamiltonian}
\setcounter{equation}{0}
\renewcommand{\theequation}{2.\arabic{equation}}

We will quickly review the Kogut-Susskind Hamiltonian formulation of
lattice gauge theory. If the reader finds this discussion
incomplete, we refer him or her to the book by Creutz \cite{Creutz}.

Consider a lattice of sites $x$ of size
$L^{1}\times L^{2}$, with
sites $x$ whose coordinates are $x^{1}$ and $x^{2}$. We require that $x^{1}/a$ and $x^{2}/a$
are integers, where
$a$ is the lattice spacing. There
are $2$ space directions, labeled
$j=1,2$. Each link is a pair $x$, $j$, and joins the site 
$x$ to $x+{\hat j}a$, where
$\hat j$ is a unit vector in the $j^{\rm th}$ direction.

We use generators 
$t_{b}$, $b=1,2,3$, of 
the Lie algebra
of SU($2$), which are related to the Pauli matrices by $t^{b}=\sigma^{b}/{\sqrt 2}$. The
identity matrix will be denoted by 
$\ident$.

For now, the Hamiltonian lattice gauge theory will be in the
temporal gauge $A_{0}=0$. The basic degrees of freedom, before 
any further 
gauge fixing, are elements of the group SU($2$) in the 
fundamental ($2\times 2$)-dimensional 
matrix representation $U_{j}(x)\in$ SU($2$) at each link
$x$, $j$. The relation between these variables and
the continuum gauge field is $U_{j}(x)=e^{-{\rm i}aA_{j}(x)}$. There 
are three self-adjoint electric-field 
operators at each link
$l_{j}(x)_{b}$, $b=1,2,3$. The 
commutation relations on the
lattice are
\begin{eqnarray}
[l_{j}(x)_{b} , l_{k}(y)_{c} ]=
{\rm i}{\sqrt 2}\delta_{x\,y}\delta_{j\,k} \;\epsilon^{bcd}
\;l_{j}(x)_{d} \;, \nonumber 
\end{eqnarray}
\begin{eqnarray}
[l_{j}(x)_{b}, U_{k}(y)]        =
-\delta_{x\,y}\delta_{j\,k}\; t_{b}\;U_{j}(x)\;,
\label{loccommrel}
\end{eqnarray}
all others zero.

The Hamiltonian is
\begin{eqnarray}
H= \sum_{x } 
\sum_{j=1}^{2} \sum_{b=1}^{3}
\;\frac{g_{0}^{2}}{2a}\left[ \, l_{j}(x)_{b} \,\right]^{2}-
\sum_{x} 
\frac{1}{2g_{0}^{2}a}\;
\left[ {\rm Tr}\; U_{1\,2}(x) +{\rm Tr}\;U_{2\,1}(x)\right]\;, \label{hamilt1}
\end{eqnarray}
where
\begin{eqnarray}
U_{j\,k}(x)  =
U_{j}(x)
U_{k}(x+{\hat j}a)
U_{j}(x+{\hat k}a)^{\dagger}
U_{k}(x)^{\dagger} \;,
\nonumber
\end{eqnarray}
where $\hat j$ and $\hat k$ are the unit vectors in the $j$- and $k$-directions, respectively, 
and the bare coupling constant $g_{0}$ is dimensionless. The coefficient
of the kinetic term is just half the square of the continuum coupling constant $e$, namely
$g_{0}^{2}/(2a)=e^{2}/2$. This is why the mass gap in $(2+1)$ dimensions is 
proportional to
$e^{2}$ in strong-coupling expansions.

We denote 
the adjoint representation of the SU($2$) gauge field by 
$\mathcal R_{j}(x)$. The precise definition is
${\mathcal R}_{j}(x)_{b}^{\;\;c}t_{c}=
U_{j}(x)t_{b}U_{j}(x)^{\dagger}$. 
The matrix ${\mathcal R}_{j}$ lies in the group  SO($3$).

Color charge operators $q(x)_{b}$, may be placed at lattice sites. These obey
\begin{eqnarray}
[q(x)_{b},q(y)_{c}]={\rm i} {\sqrt{2}}\epsilon^{bca} \delta_{xy}q(x)_{a} \; . \label{charge-comm}
\end{eqnarray}
In the presence of static charges, Gauss's law is the condition on physical wave functions
\begin{eqnarray}
\left[( {\cal D} \cdot l)(x)_{b}-q(x)_{b}\right]\; \Psi( \{U \}) =0 \;.
\label{gauss-charge}
\end{eqnarray}
where 
\begin{eqnarray}
\left[{\cal D}_{j} l_{j}(x)\right]_{b}  = 
l_{j}(x)_{b}-
{\cal R}_{j}(x-{\hat j}a )_{b}^{\;\;\;c} \;l_{j}(x -{\hat j}a )_{c} \;. \label{cov-deriv}
\end{eqnarray}

\section{The axial gauge; splitting the Hamiltonian}
\setcounter{equation}{0}
\renewcommand{\theequation}{3.\arabic{equation}}

Next we discuss the axial-gauge-fixing procedure. This 
is most easily done on a cylinder
of dimensions $L^{1}\times L^{2}$, with open boundary conditions in the $x^{1}$-direction and
periodic boundary conditions in the $x^{2}$-direction \cite{PhysRevD71}. For any function 
$f(x^{1},x^{2})$, we require that $f(x^{1},x^{2}+L^{2})=f(x^{1},x^{2})$. The range
of coordinates is $x^{1}=0,a,2a,\dots,L^{1}$, $x^{2}=0,a,2a,\dots, L^{2}-a$. For any physical
state $\Psi$, Gauss's law implies that 
\beq
l_{1}(x^{1},x^{2})\Psi =\left[{\mathcal R}_{1}(x^{1}-a,x^{2})l_{1}(x^{1}-a,x^{2})-
({\mathcal D}_{2}\, l_{2}) (x^{1}-a, x^{2}) +q(x) \right] \Psi \;,  \nonumber
\eeq
so the operators on each side of this expression may be identified. Taking
the gauge condition $U_{1}(x^{1},x^{2})=\ident$, which is possible everywhere on
a cylinder, we sum over the $1$-coordinate to obtain
\beq
l_{1}(x^{1},x^{2})=
\sum_{y^{1}=0}^{x^{1}}
q(y^{1},x^{2})
-\sum_{y^{1}=0}^{x^{1}}({\mathcal D}_{2}\,l_{2}) (y^{1}, x^{2}) \;,
\label{lattice-electric}
\eeq
which is the lattice analogue of (\ref{continuum-electric}). There is a remnant of
Gauss's law not determined by (\ref{lattice-electric}), which is the condition
\beq
\left[
\sum_{x^{1}=0}^{L^{1}}({\mathcal D}_{2}\,l_{2})(x^{1},x^{2})-\sum_{x^{1}=0}^{L^{1}}q(x^{1},x^{2})
\right] \Psi=0 \label{lattice-physical}
\eeq
on physical states $\Psi$. The condition (\ref{lattice-physical}) is the lattice analogue
of (\ref{continuum-physical}).

In the axial gauge, using the nonlocal expression (\ref{lattice-electric})
for the electric field in the $x^{1}$-direction
(which henceforth will be called the horizontal direction), the Hamiltonian 
(\ref{hamilt1}) becomes $H=H_{0}+H_{1}$, where 
\beq
H_{0}=\sum_{x^{2}=0}^{L^{2}-a} H_{0}(x^{2})\;, \label{sum-of-two-terms}
\eeq
with
\beq
H_{0}(x^{2})= \sum_{x^{1}=0 }^{L^{1}} 
\;\frac{g_{0}^{2}}{2a}[ l_{2}(x^{1},x^{2})]^{2}-
\sum_{x=0}^{L^{1}-a} 
\frac{1}{2g_{0}^{2}a}\;
\left[ {\rm Tr}\; U_{2}(x)^{\dagger}U_{2}(x^{1}+a,x^{2}) +c.c.\right] \;, \label{lattice-sigma}
\end{eqnarray}
and
\beq
H_{1}&=&-\frac{(g_{0}^{\prime})^{2}}{2a}\;\sum_{x^{2}=0}^{L^{2}-a} \;\sum_{x^{1},y^{1}=0}^{L^{1}}
{\vert x^{1}-y^{1}\vert } \nonumber \\
&\times&\!\! \left[ l_{2}(x^{1},x^{2})-{\mathcal R}_{2}(x^{1},x^{2}-a)l_{2}(x^{1},x^{2}-a) -q(x^{1},x^{2})   \right]   \nonumber  \\
&\times& \!\!\left[ l_{2}(y^{1},x^{2})-{\mathcal R}_{2}(y^{1},x^{2}-a)l_{2}(y^{1},x^{2}-a)  
-q(y^{1},x^{2})    \right] \;,
\label{nonlocal}
\eeq
and where we have now introduced the second dimensionless coupling 
constant $g_{0}^{\prime}$, defined by $(e^{\prime})^{2}=(g_{0}^{\prime})^{2}/a$. The only
interaction in the $x^{2}$-direction (which from now on will be called the vertical direction)
is due to $H_{1}$. 

If $g_{0}^{\prime}$ vanishes, the cylindrical lattice splits apart as shown in Figure 1. The dashed
line at the top of the unsplit lattice on the left in this figure
indicates that this line is identified with the
bottom line. No such identification is made in the split lattice on the right.

\vspace{5pt}
\begin{center}
\begin{picture}(150,80)(0,0)

\multiput(13,15)(0,5){9}{\multiput(0,0)(5,0){10}{\put(0,0){\line(1,0){5}}
\put(0,0){\line(0,1){5}}
\put(0,5){\line(1,0){5}}
\put(5,0){\line(0,1){5}}}}
\multiput(13,60)(5,0){10}{\put(0,0){\line(1,0){5}}
\put(0,0){\line(0,1){5}}
\put(0,4.5){-} \put(1.7,4.5){-} \put(3.3,4.5){-}
\put(5,0){\line(0,1){5}}}

\multiput(88,3)(0,8){10}{\multiput(0,0)(5,0){10}{\put(0,0){\line(1,0){5}}
\put(0,0){\line(0,1){5}}
\put(0,5){\line(1,0){5}}
\put(5,0){\line(0,1){5}}}}

\put(73,40){$\longrightarrow$}

\put(11,8){$0$}
\put(16,8){$a$}
\put(21,8){$2a$}
\multiput(28,8)(5,0){7}{$\cdot$}
\put(61,8){$L^{1}$}

\put(13,2){\vector(1,0){50}}
\put(66,1){$x^{1}$}

\put(8,14){$0$}
\put(8,19){$a$}
\put(7,24){$2a$}
\multiput(8,29)(0,5){7}{$\cdot$}
\put(8,65){$L^{2}$}

\put(1,15){\vector(0,1){50}}
\put(0,69){$x^{2}$}

\end{picture}\\

\end{center}

\vspace{5pt}

\begin{center}

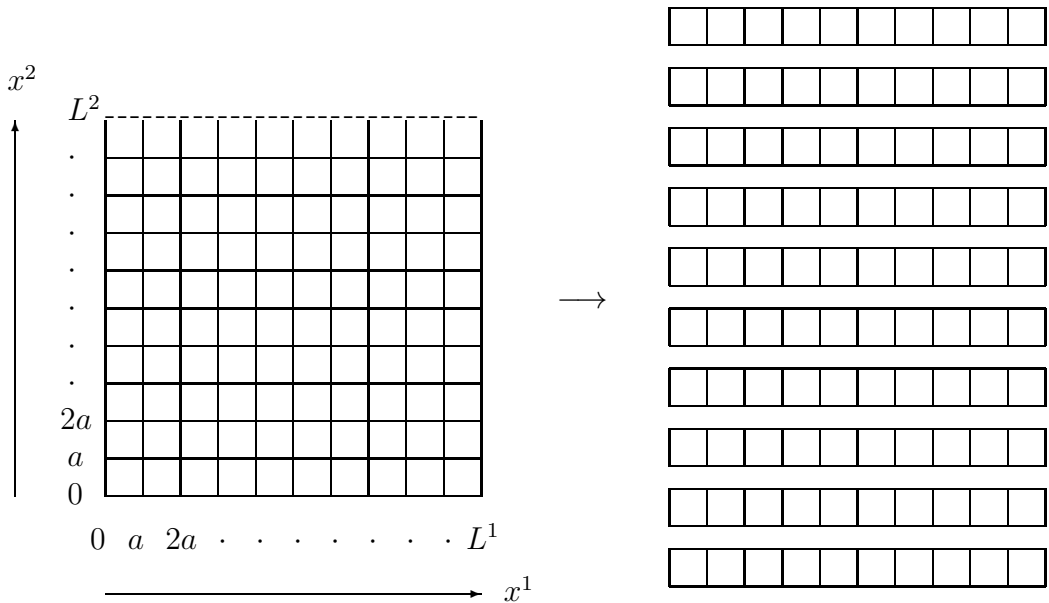
{\bf Figure 1. The splitting of the lattice at $g_{0}^{\prime}= 0$.}

\end{center}

The operators $H_{0}(x^{2})$ are Hamiltonians of ${\rm SU}(2)\!\times\!{\rm SU}(2)$ principal chiral nonlinear
sigma models, as noted in reference \cite{PhysRevD71} (see also reference \cite{Griffin}). Each sigma model is represented by a horizontal ladder of plaquettes on the right-hand side of
Figure 1. We see that there
is a ladder for each value of $x^{2}$. Setting $g_{0}^{\prime}$ to zero results
in decoupled layers of $(1+1)$-dimensional sigma models. Increasing  $g_{0}^{\prime}$
leads to an interaction between the vertically-separated layers. This
fact was used to give a set of 
simple arguments for confinement for $g_{0}^{\prime}\ll g_{0}$. The expression for $H_{1}$, acting
on physical states, with $q=0$, is 
identical to the expression on the right-hand side of equation (3.6) in reference
\cite{PhysRevD71}, by virtue of (\ref{lattice-physical}).

For readers who are not familiar with Hamiltonian strong-coupling expansions, we remark that
they start by neglecting the second term of (\ref{hamilt1}) or (\ref{lattice-sigma}). This term is
reintroduced by Rayleigh-Schr\"{o}dinger perturbation theory, yielding the strong coupling
expansion in $1/g_{0}^{2}$.

In a strong-coupling expansion, the magnetic flux is allowed to flow through space 
unconstrained. By including the ``magnetic" or ``plaquette" term of order $1/g_{0}^{2}$, perturbatively,
this unphysical assumption is corrected for. In contrast, we start by allowing 
the $1$-component of the
electric field to flow through space 
unconstrained. By including $H_{1}$ (we shall discuss how, at the end of
the next section) we correct for this assumption.

At this point, the reader has noticed that  our aim is to exploit a type of dimensional reduction
from $(2+1)$ to $(1+1)$ dimensions. This reduction is very different from ``compactification",
that is, making $L^{2}$ small. It is in some sense the opposite of
Fu and Nielsen's idea of a ``layer phase" \cite{FuNielsen}, in which the coupling between lattice
layers is made strong instead of weak. The reduction is similar to the
``deconstruction" of 
Arkani-Hamed, Cohen and Georgi \cite{deconstruction} in that the difference between the $(1+1)$-dimensional
Hamiltonians and the $(2+1)$-dimensional Hamiltonians in that the sigma models can be 
regarded as gauged and coupled together through the external gauge field.

\section{The effective action for the electrostatic potential}
\setcounter{equation}{0}
\renewcommand{\theequation}{4.\arabic{equation}}

Before we can make much use of the axial-gauge formulation, we need to examine
$H_{0}$ and $H_{1}$ in more detail. First we consider $H_{0}$. If we adopt the interaction
representation, time derivatives of operators $\mathfrak A$ are given by
$\partial_{0}{\mathfrak A}={\rm i}[H_{0},{\mathfrak A}]$. By working out the time derivative
of $U_{2}(x^{1},x^{2})$, we find
\beq
l(x^{1},x^{2})_{b}\!\!&\!\!=\!\!&\!\!
\frac{{\rm i}a}{g_{0}^{2}}\, {\rm Tr}\, t_{b}\, \partial_{0} {U}(x^{1},x^{2})\,
U(x^{1},x^{2})^{\dagger}\;,    \nonumber  \\
{\mathcal R}(x^{1},x^{2})_{b}^{\;\;\;c}\,l(x^{1},x^{2})_{c}\!\!&\!\!=\!\!&\!\!
\frac{{\rm i}a}{g_{0}^{2}}\, {\rm Tr}\, t_{b}\,U(x^{1},x^{2})^{\dagger} \,
\partial_{0} {U}(x^{1},x^{2})\;,    \label{interaction-electric-field}
\eeq
where we have dropped the subscript $2$, since there is only one spatial component of
the gauge field. The time dependence of operators is implicit in these expressions. The 
$(1+1)$-dimensional ${\rm SU}(2)
\!\times\! {\rm SU}(2)$ principal chiral sigma model of the field $U\in {\rm SU}(2)$ has the Lagrangian
\beq
{\mathcal L}_{\rm PCSM}=\frac{1}{2g_{0}^{2}}\,\eta^{\mu \nu}\,{\rm Tr} \,\partial_{\mu}
U^{\dagger}\partial_{\nu}U
\;, \;\;\mu,\nu=0,1\;.
\label{lagrangian}
\eeq
The left-handed and right-handed currents are, respectively,
\beq
j^{\rm L}_{\mu}(x)_{b}={\rm i}{\rm Tr}\,t_{b} \, \partial_{\mu}U(x)\, U(x)^{\dagger}\;,\;\;
j^{\rm R}_{\mu}(x)_{b}={\rm i}{\rm Tr}\,t_{b} \, U(x)^{\dagger}\partial_{\mu}U(x) \;. \label{current-definition}
\eeq
The Hamiltonian obtained from (\ref{lagrangian}) is
\beq
H_{\rm PCSM}\!=\!\int dx^{1} \frac{1}{2g_{0}^{2}}\{ [j^{\rm L}_{0}(x)_{b}]^{2}+[j^{\rm L}_{1}(x)_{b}]^{2}\}
\!=\!\int dx^{1} \frac{1}{2g_{0}^{2}}\{ [j^{\rm R}_{0}(x)_{b}]^{2}+[j^{\rm R}_{1}(x)_{b}]^{2}\}. \label{HNLSM}
\eeq
By comparing (\ref{interaction-electric-field}) with 
(\ref{current-definition}), we can see that $H_{\rm PCSM}$ in (\ref{HNLSM})
is the continuum limit of $H_{0}(x^{2})$ in (\ref{lattice-sigma}).

Next we turn to $H_{1}$. This interaction Hamiltonian is nonlocal, but can be made local
by reintroducing the temporal component of the gauge field. In one continuous infinite dimension
$\mathbb R$, the function
$g(x^{1}-y^{1})=\vert x^{1}-y^{1} \vert/2$ is the Green's function of the ``Laplacian"; in other
words
\beq
-\partial_{1}^{2}\,g(x^{1}-y^{1})=\delta(x^{1}-y^{1}) \;. \nonumber
\eeq 
On our lattice, with $x^{1}$ and $y^{1}$ taking values $0, \;a,\;2a,\;\dots,\;L^{1}$, the
same function $g(x^{1}-y^{1})=\vert x^{1}-y^{1} \vert/2$ is the Green's function of an $(L^{1}/a+1)$-dimensional operator
$\Delta_{L^{1},a}$, by which it is meant
\beq
\Delta_{L^{1},a}\,g(x^{1}-y^{1})=
\sum_{z^{1}=0}^{L^{1}}\left( \Delta_{L^{1},a}\right)_{x^{1} z^{1}}\,g(z^{1}-y^{1})=
\frac{1}{a}\delta_{x^{1}y^{1}} \;. \nonumber
\eeq 
In the continuum limit $a\rightarrow 0$ and thermodynamic 
limit $L^{1}\rightarrow \infty$, $\Delta_{L^{1},a}\rightarrow -\partial_{1}^{2}$. We use this operator to introduce
an auxiliary field $\Phi(x^{1},x^{2})_{b}$ to replace (\ref{nonlocal}) by
\beq
H_{1}&=& \sum_{x^{2}=0}^{L^{2}-a}  \sum_{x^{1}=0}^{L^{2}} \left\{
\frac{(g_{0}^{\prime})^{2}a}{4}\,\Phi(x^{1},x^{2})\Delta_{L^{1},a}\Phi(x^{1},x^{2}) \right. \nonumber \\
&-\!\!&\!\! (g_{0}^{\prime})^{2}
\left[ l_{2}(x^{1},x^{2})-{\mathcal R}_{2}(x^{1},x^{2}-a)l_{2}(x^{1},x^{2}-a) \right. \nonumber \\ 
&-&q(x^{1},x^{2})   \left]  
\Phi(x^{1},x^{2})  \right\} .
\label{local}
\eeq
Let us assume that there are only two color charges - a quark with charge $q$ at site $u$
and another quark with charge $q^{\prime}$ at site $v$ (note: the gauge group is SU($2$), so it
makes no difference if we have a pair of heavy quarks or a heavy quark and antiquark). For small lattice spacing, we approximate the sum over $x^{1}$ {\it only} as an integral to obtain
\beq
H_{1}\!\!&\!\!=\!\!&\!\! \sum_{x^{2}=0}^{L^{2}-a}  \int dx^{1} \,
\frac{(g_{0}^{\prime})^{2}a^{2}}{4}\,\partial_{1}\Phi(x^{1},x^{2})\partial_{1}\Phi(x^{1},x^{2}) \nonumber \\
\!\!&\!\!-\!\!&\!\! 
\left(\frac{g_{0}^{\prime}}{g_{0}}\right)^{2}\,\,\sum_{x^{2}=0}^{L^{2}-a}  \int dx^{1} \!\!
\left[ j^{\rm L}_{0}(x^{1},x^{2})\Phi(x^{1},x^{2}) -j^{\rm R}_{0}(x^{1},x^{2}) \Phi(x^{1},x^{2}+a) \right]  
\nonumber \\
&+&(g_{0}^{\prime})^{2}q_{b}\Phi(u^{1},u^{2})_{b} -(g_{0}^{\prime})^{2}
q^{\prime}_{b}\Phi(v^{1},v^{2})_{b}   \; .
\label{continuum-local}
\eeq
We wish to stress that we are not really taking $a\rightarrow 0$, but only assuming that $a$ is
small. Though the expression (\ref{local}) makes the physical meaning of the interaction
clearer than (\ref{local}), we shall keep the regulator, at least implicitly. From 
the coupling to charges, we see that $\Phi_{b}$ is proportional to the temporal component
of the gauge field ${A_{0}}_{b}$. We will call $\Phi_{b}$ the electrostatic potential for this reason.

From (\ref{continuum-local}) we see that the left-handed charge of the sigma model
at $x^{2}$ is coupled to the electrostatic potential at $x^{2}$. The right-handed charge
of the sigma model is coupled to the electrostatic potential at $x^{2}+a$.

We now state the mathematical problem we wish to solve. In the presence of a quark at
$u$ and an antiquark at $v$, what is the effective action ${\mathfrak S}(\Phi)$, after 
integrating out
$U$? If $u^{2}=v^{2}$, that is, the quarks are only separated horizontally, the effective action
is given by
\beq
e^{{\rm i}{\mathfrak S}(\Phi)}=\left< 0 \right\vert  {\mathcal T} e^{-i\int dx^{0} H_{1}} \left\vert 0 \right>\;,
\label{effective-action}
\eeq
where the state $\left\vert 0 \right>$ is the tensor product of sigma-model vacua (in other
words, it is the vacuum of $H_{0}$) and
where $\mathcal T$ denotes time ($x^{0}$) ordering. If $u^{2}\neq v^{2}$, the expression
(\ref{effective-action}) is no longer correct. In that case, the expectation value needs to be
taken with respect to different eigenstates of $H_{0}$, as we shall discuss in Section 7.

The effective action may be expanded in terms of vacuum expectation values
of products of currents:
\beq
{\rm i}\!\!\!&\!\!\!{\mathfrak S}\!\!\!&\!\!\!(\Phi)= -{\rm i}\sum_{x^{2}=0}^{L^{2}-a} \int d^{2}x \,
(g_{0}^{\prime})^{2}a^{2}\,\partial_{1}\Phi(x^{0},x^{1},x^{2})\partial_{1}\Phi(x^{0},x^{1},x^{2}) \nonumber \\
\!\!&\!\!-\!\!&\!\!\frac{1}{2}\left(\frac{g_{0}^{\prime}}{g_{0}}\right)^{4}\,\,\sum_{x^{2}=0}^{L^{2}-a}\!\int d^{2}x \!\!\int d^{2}y 
\left[
\left<0\right\vert {\mathcal T}\;j^{\rm L}_{0}(x^{0},x^{1},x^{2})_{b}\; j^{\rm L}_{0}(y^{0},y^{1},x^{2})_{c}
\left\vert 0\right> \right. \nonumber \\
\!\!&\!\!+\!\!&\;\left. \left<0\right\vert {\mathcal T}\;j^{\rm R}_{0}(x^{0},x^{1},x^{2})_{b}\; 
j^{\rm R}_{0}(y^{0},y^{1},x^{2})_{c}
\left\vert 0\right> \right] \Phi(x^{0},x^{1},x^{2})_{b} \Phi(y^{0},y^{1},x^{2})_{c}  \;+O(\Phi^{4}) \nonumber \\
&-&{\rm i}\int dx^{0}\left[ (g_{0}^{\prime})^{2}q(x^{0})_{b}\Phi(x^{0},u^{1},u^{2})_{b} -(g_{0}^{\prime})^{2}
q^{\prime}(x^{0})_{b}\Phi(x^{0},v^{1},u^{2})_{b}\right]  \nonumber \\
&\!\!+\!\!&{\rm i} S_{\rm WZWN}(q)+{\rm i}S_{\rm WZWN}(q^{\prime})  \; ,
\label{leading-order}
\eeq
where $d^{2}x=dx^{0}dx^{1}, d^{2}y=dy^{0}dy^{1}$ and $S_{\rm WZWN}(q)$ is the Wess-Zumino-Witten-Novikov action of a single SU($2$) quark charge (see for example references \cite{bal}) the details of which are not important for our purposes. We will determine the two-point correlators of currents
in (\ref{leading-order}) at large separations. 

If we ignore the quantum corrections in (\ref{leading-order}), the potential between the quark and the
antiquark is 
\beq 
V(u^{1}-v^{1})=\sigma \vert u^{1}-v^{1}\vert\;,\;\;\sigma=
q^{2}\frac{(g_{0}^{\prime})^{4}}{(g_{0}^{\prime})^{2}a^{2}}=
\frac{3}{2}\frac{(g_{0}^{\prime})^{2}}{a^{2}}\;. \nonumber
\eeq
This is the
result of Section 6 of reference \cite{PhysRevD71}. In the next section, we show the quantum corrections
from the current-current correlators will drastically change this result. Physically, these correlators 
correspond to transverse (that is, vertical) fluctuations of 
the electric string, as we discuss in Section 7.

\section{The leading-order corrections to the effective action}
\setcounter{equation}{0}
\renewcommand{\theequation}{5.\arabic{equation}}

From the exact result for the two-point form factor (\ref{current-form-factor}), discussed in
the appendix, we will
determine the correlation functions 
\beq
D(x,y)_{bc}\!\!&\!\!=\!\!&\!\!
\left<0\right\vert {\mathcal T}\;j^{\rm L}_{0}(x^{0},x^{1},x^{2})_{b}\; j^{\rm L}_{0}(y^{0},y^{1},x^{2})_{c}
\left\vert 0\right>  \nonumber \\
\!\!&\!\!=\!\!&\!\!
\left<0\right\vert {\mathcal T}\;j^{\rm R}_{0}(x^{0},x^{1},x^{2})_{b}\; j^{\rm R}_{0}(y^{0},y^{1},x^{2})_{c}
\left\vert 0\right> \;, \label{current-correlation-functions}
\eeq
in (\ref{leading-order}) for the ${\rm SU}(2)\! \times \!{\rm SU}(2)\simeq {\rm O}(4)$ sigma model. We will then use this result to find the string
tension for horizontally-separated color charges. 

The {\em exact} correlation functions will also
have contributions from two- as well as all higher-point form factors. The 
complete formula for the 
Wightman (non-time-ordered)
expectation value of two operators in terms of form factors is
\beq
\left< 0 \right\vert \!\!\!&\!\!\!
{\mathfrak B}\!\!\!&\!\!\!(x)  {\mathfrak C}(y) \left\vert 0\right>=
\left< 0 \right\vert
{\mathfrak B}(x) \left\vert 0 \right> \! \left< 0 \right\vert  {\mathfrak C}(y) \left\vert 0\right> \nonumber \\
\!\!&\!\!+\!\!&\!\!\sum_{M=1}^{\infty} \int\frac{d\theta_{1}\cdots d\theta_{M}}{(2\pi)^{M}M!}
\left< 0 \right\vert
{\mathfrak B}(x) \left\vert \theta_{M},j_{M}\dots,\theta_{1},j_{1} \right>   \!\!
\left< \theta_{1},j_{1},\dots,\theta_{M},j_{M} 
\right\vert  {\mathfrak C}(y) \left\vert 0\right>   \!.
\label{decomposition}
\eeq
To obtain this result, we used the resolution of the identity (\ref{resolution}). Time-ordered expectation values can be written in terms of Wightman functions. In a field theory
with a mass gap $m$, the largest contribution to (\ref{decomposition}) for large $m\vert x-y\vert$
comes from the terms with the smallest number of particle exchanges $M$. For our problem, with
two-sigma model charge densities, the first
term, {\em i.e.} the vacuum channel, gives no contribution; therefore we can consider just the
case of $M=2$. We will evaluate
(\ref{current-correlation-functions}) in this way. Viewing the two-point form
factor as a vertex between the electrostatic potential $\Phi$ and two excitations of the
sigma model, called Faddeev-Zamolodchikov or FZ particles, we consider the one-loop diagram:

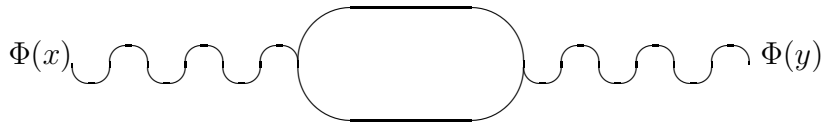
\begin{center}

\begin{picture}(150,20)(-10,0)

\put(64.5,10){\oval(30,15)}

\put(22,10){\oval(5,5)[b]}
\put(27,10){\oval(5,5)[t]}
\put(32,10){\oval(5,5)[b]}
\put(37,10){\oval(5,5)[t]}
\put(42,10){\oval(5,5)[b]}
\put(47,10){\oval(5,5)[t]}

\put(82,10){\oval(5,5)[b]}
\put(87,10){\oval(5,5)[t]}
\put(92,10){\oval(5,5)[b]}
\put(97,10){\oval(5,5)[t]}
\put(102,10){\oval(5,5)[b]}
\put(107,10){\oval(5,5)[t]}

\put(11,10){$\Phi(x)$}
\put(111,10){$\Phi(y)$}

\end{picture}

\end{center}

\noindent
This diagram is infrared and ultraviolet finite. A scale is set by the sigma-model mass gap $m$. We
will expand this diagram in derivatives of $\Phi$. Lowest order in derivatives means lowest
order in momentum, in the Fourier transform of this amplitude. The two-particle form factors should
therefore be sufficient. If we wanted 
many higher-derivative terms in the effective action of the electrostatic potential 
${\mathfrak S}(\Phi)$, this would no longer be the case.

Taking the complex conjugate of the expression for the form factor
(\ref{useful-form-factor}), in the appendix, with $x$ replaced by $y$,
applying (\ref{decomposition}), truncating $M>2$, and finally ordering in the time
coordinates $x^{0}$ and
$y^{0}$, yields the following for the current-current correlation functions (\ref{current-correlation-functions})
\beq
D(\!\!\!\!&\!\!\!x\!\!\!&\!\!\!,y)_{bc}=\frac{4m^{2}\delta_{bc} }{2! (2\pi)^{2}} \int d \theta_{1} d \theta_{2}\,
(\cosh\theta_{1}-\cosh\theta_{2})^{2} \,\left\vert F(\theta_{2}-\theta_{1}) \right\vert^{2} \nonumber \\
\!\!&\!\! \times \!\!&\!\! \exp \left\{ -{\rm i} m \,
{\rm sgn}(x^{0}-y^{0})\,[ (x^{0}-y^{0})(\cosh\theta_{1}+\cosh \theta_{2}) \right.  \nonumber \\
\!\!&\!\!-\!\!&\!\!(x^{1}-y^{1}) \left.
(\sinh\theta_{1}+\sinh \theta_{2})
] \right\} \;, \label{preliminary-one-loop}
\eeq
where sgn$(x^{0}-y^{0})$ is defined as $1$ if $x^{0}>y^{0}$ and $-1$ if $x^{0}<y^{0}$.

The integration in (\ref{preliminary-one-loop}) can be made somewhat easier by introducing
new parameters  $\Omega=(\theta_{1}+\theta_{2})/2$ and $\omega=(\theta_{1}-\theta_{2})/2$:
\beq
D(\!\!\!\!&\!\!\!x\!\!\!&\!\!\!,y)_{bc}=\frac{4m^{2}\delta_{bc} }{\pi^{2}} \int d \Omega d \omega\,
\sinh^{2}\Omega \, \sinh^{2} \omega  \nonumber \\
\!\!&\!\! \times \!\!&\!\! \exp\{-2{\rm i} m \, \cosh\Omega\,
{\rm sgn}(x^{0}-y^{0})\,[ (x^{0}-y^{0})\cosh\omega 
-(x^{1}-y^{1})
\sinh\omega
] \}        \nonumber \\
\!\!&\!\! \times \!\!&\!\!  \exp-\int_{0}^{\infty} \frac{d\xi}{\xi} \frac{e^{-\xi}}{\cosh^{2}\frac{\xi}{2}}
\left(1-\cosh\xi \cos \frac{2\xi \omega}{\pi} \right)
\;. \label{preliminary-one-loop1}
\eeq

The action ${\mathfrak S}(\Phi)$ is nonlocal, but the excitations of the sigma 
model are massive particles. Hence we expect that ${\mathfrak S}(\Phi)$
is dominated by 
local terms, obtained from the derivative expansion. Let us introduce the new coordinates
$X^{\mu}$ and $r^{\mu}$ by $x^{\mu}=X^{\mu}+r^{\mu}/2$ and $y^{\mu}=X^{\mu}-r^{\mu}/2$. We
expand $\Phi(x)$ and $\Phi(y)$ in powers of $r^{\mu}$, in the standard way as
\beq
\Phi(x)\!\!&\!\!=\!\!&\!\!\Phi(X)+
\frac{r^{\mu}}{2}\partial_{\mu}\Phi(X)+\frac{r^{\mu}r^{\nu}}{8}\partial_{\mu}\partial_{\nu}
\Phi(X)+\cdots\;, \nonumber \\
\Phi(y)\!\!&\!\!=\!\!&\!\!\Phi(X)-
\frac{r^{\mu}}{2}\partial_{\mu}\Phi(X)+\frac{r^{\mu}r^{\nu}}{8}\partial_{\mu}\partial_{\nu}
\Phi(X) \pm \cdots\;, \label{der-exp}
\eeq
where $\partial_{\mu}$ now denotes $\partial/\partial X^{\mu}$.

Taking care to sum over $L$ and $R$, the term we want to evaluate in the effective
action, which is quadratic in the fields, is given by
\beq
{\rm i} {\mathfrak S}^{(2)}(\Phi)\!\!&\!\!=\!\!&\!\!-\left(\frac{g_{0}^{\prime}}{g_{0}}\right)^{4}\,\,
\sum_{x^{2}=0}^{L^{2}-a} \int d^{2}X\, d^{2}r\,
D\!\left(X+\frac{r}{2}, X-\frac{r}{2}\right)_{bc}    \nonumber \\
\!\!&\!\! \times \!\!&\!\!
\Phi\!\left(X+\frac{r}{2},x^{2}\right)_{b}\!\Phi\!\left(X-\frac{r}{2}, x^{2}\right)_{c} \!. 
\label{action-correction}
\eeq

At the risk of overemphasizing a point, we 
remark that the truncation of (\ref{decomposition}) 
to $M=2$ and the use of the derivative
expansion (\ref{der-exp}) have the same justification. Both are valid approximations
in a massive theory. Though we expand in $r$ in (\ref{der-exp}), it is not a short-distance
expansion in the usual sense. We can integrate over $r$ precisely because large
$m\vert r\vert$ contributions are suppressed. The result of substituting (\ref{der-exp}) into
(\ref{action-correction}) is really a small-momentum expansion.

After substituting (\ref{preliminary-one-loop1}) and (\ref{der-exp}) into (\ref{action-correction})
we integrate over $r^{\mu}$. The integrals over $r^{\mu}$ are are of the form
\beq
I(Q_{0},Q_{1}, A)=\int d^{2}r\,e^{-{\rm i}Q_{0}\vert r^{0}\vert +iQ_{1}r^{1}{\rm sgn(r^{0})}}
A(r^{0},r^{1})\;,  \nonumber
\eeq
for some polynomial $A(r^{0},r^{1})$, where
\beq
Q_{0}= 2m\cosh\Omega \cosh \omega\;,\;\;Q_{1}=2m\cosh\Omega \sinh \omega\;. \nonumber
\eeq
We therefore need to evaluate $I(Q_{0},Q_{1}, A)$, for
a few choices of $A(r^{0},r^{1})$. 

Suppose that $A(r^{0},r^{1})$ has no dependence on $r^{1}$. Then the $r^{1}$-integration
will produce a term proportional to $\delta(Q_{1})$, which, in turn, is proportional
to $\delta(\sinh \omega)$. From the factor $\sinh^{2}\omega$ in
the integral expression (\ref{preliminary-one-loop1}), this will give no contribution to 
${\rm i} {\mathfrak S}^{(1)}(\Phi)$ in (\ref{action-correction}). Since we are working to quadratic
order in $r^{0}$ and $r^{1}$, we therefore need only consider 
$A(r^{0},r^{1})=r^{1}, r^{0}r^{1},(r^{1})^{2}$. 

It is elementary to show that 
$I(Q_{0},Q_{1},r^{1})=0$. 

The integral $I(Q_{0},Q_{1},r^{0}r^{1})$ is proportional to
$\delta^{\prime}(\sinh \omega)$. Again, the factor $\sinh^{2}\omega$ in
the integral expression (\ref{preliminary-one-loop1}) insures that this will give no contribution to 
${\rm i} {\mathfrak S}^{(2)}(\Phi)$ in (\ref{action-correction}).

The only integral remaining is 
\beq
I\!\left[Q_{0},Q_{1},(r^{1})^{2}\right]=-\frac{2{\rm i}\delta^{\prime \prime}(Q_{1})}{Q_{0}-{i}\varepsilon}
=-{\rm P.V.}\frac{2{\rm i}\delta^{\prime \prime}(Q_{1})}{Q_{0}}
+2\pi\delta(Q_{0})\delta^{\prime \prime}(Q_{1})
\;. \nonumber
\eeq
Only the principal value contributes to the effective action, which is 
\beq
{\rm i} {\mathfrak S}^{(2)}(\Phi)\!\!&\!\!=\!\!&\!\!-\frac{2m{\rm i}}{\pi^{2}}
\left(\frac{g_{0}^{\prime}}{g_{0}}\right)^{4}\,\,\sum_{x^{2}=0}^{L^{2}-a} \int d^{2}X
\int d\Omega d\omega \frac{\sinh^{2}\Omega \sinh^{2}\omega}{\cosh \Omega \cosh\omega}
\delta^{\prime \prime}(2m\cosh\Omega \sinh\omega)   \nonumber \\
\!\!&\!\! \times \!\!&\!\! \exp-\int_{0}^{\infty} \frac{d\xi}{\xi} \frac{e^{-\xi}}{\cosh^{2}\frac{\xi}{2}}
\left(1-\cosh\xi \cos \frac{2\xi \omega}{\pi} \right) \left[\partial_{1}\Phi(X,x^{2})_{b}\right]^{2}  \nonumber \\
&+&{\rm \;higher\;derivative\;terms}
\;.\nonumber
\eeq
The integration over $\Omega$ and $\omega$ can now be done. After some work, we find
\beq
{\rm i} {\mathfrak S}^{(2)}(\Phi)=\!\!&\!\!-\!\!&\!\!\frac{\rm i}{3m^{2}\pi^{2}}
\left(\frac{g_{0}^{\prime}}{g_{0}}\right)^{4}
\exp\left[-2\int_{0}^{\infty} \frac{d\xi}{\xi} {e^{-\xi}}{\tanh^{2}\frac{\xi}{2}}\right] 
\nonumber \\
&\!\! \times \!\!&\!\! 
\sum_{x^{2}=0}^{L^{2}-a}
\int d^{2}X \left[\partial_{1}\Phi(X,x^{2})_{b}\right]^{2}\;+\;{\rm h.\;d.\;t.}  , \label{central-result}
\eeq
which is the central result of this paper.

The correction to ${\mathfrak S}(\Phi)$ which is cubic in $\Phi$ can be shown
to vanish by symmetry considerations, ${\mathfrak S}^{(3)}(\Phi)=0$. The quartic 
correction ${\mathfrak S}^{(4)}(\Phi)$, does not vanish. This can also 
be obtained using form-factor methods, though
the calculation will be longer than that above. An interesting aspect of 
${\mathfrak S}^{(4)}(\Phi)$ is that it couples fields together at neighboring values of the
vertical coordinate, {\em e.g.} $x^{2}$ and $x^{2}+a$. It may lead to interesting
nonlinear dynamics of electric strings. At this order in $g_{0}^{\prime}$ the form factors and
the mass spectrum will be altered \cite{Delfino}, which will also need to be implemented.

\section{The horizontal string tension}
\setcounter{equation}{0}
\renewcommand{\theequation}{6.\arabic{equation}}

With our result (\ref{central-result}) serving as the second term of the expression (\ref{leading-order})
for the effective action ${\mathfrak S}(\Phi)$, we find
\beq
\!\!\!&\!\!\!{\mathfrak S}\!\!\!&\!\!\!(\Phi)=-
K\sum_{x^{2}=0}^{L^{2}-a}
\int d^{2} x  \left[\partial_{1}\Phi(x,x^{2})_{b}\right]^{2}  + {\rm h.\;d.\; t.}+O(\Phi^{4})   \nonumber \\
&-&\int dx^{0}\left[ (g_{0}^{\prime})^{2}q(x^{0})_{b}\Phi(x^{0},u^{1},u^{2})_{b} -(g_{0}^{\prime})^{2}
q^{\prime}(x^{0})_{b}\Phi(x^{0},v^{1},u^{2})_{b}\right]  \nonumber \\
&\!\!+\!\!&S_{\rm WZWN}(q)+S_{\rm WZWN}(q^{\prime})  \; ,
\label{the-effective-action!}
\eeq
Where the factor $K$ is given by
\beq
K=\frac{(g_{0}^{\prime})^{2}a^{2}}{4}+
\frac{1}{3m^{2}\pi^{2}}\left(\frac{g_{0}^{\prime}}{g_{0}}\right)^{4}
\exp\left[-2\int_{0}^{\infty} \frac{d\xi}{\xi} {e^{-\xi}}{\tanh^{2}\frac{\xi}{2}}\right]  \;. 
\label{constant-of-field-squared}
\eeq
Notice that there is no induced mass term in $\Phi$. This is not hard to understand; it is
due to the 
fact that the FZ particles of the principal chiral sigma model are adjoint 
charges, hence do not screen
quarks. 

Notice that at leading order, there are no time derivatives 
of $\Phi$ in ${\mathfrak S}(\Phi)$. The effective Hamiltonian is
\beq
E&=&K\sum_{x^{2}=0}^{L^{2}-a}
\int d x^{1}  \left[\partial_{1}\Phi(x,x^{2})_{b}\right]^{2}  + {\rm h.\;d.\; t.}+O(\Phi^{4})   \nonumber \\
&+\!\!&\!\! (g_{0}^{\prime})^{2}q_{b}\Phi(u^{1},u^{2})_{b} -(g_{0}^{\prime})^{2}
q^{\prime}_{b}\Phi(v^{1},u^{2})_{b}\;. \label{effective-hamiltonian}
\eeq
Thus the horizontal string tension is 
\beq
\sigma_{\rm H}= \frac{1}{4} (q_{b})^{2} \frac{(g_{0}^{\prime})^{4}}{K}=
\frac{3 (g_{0}^{\prime})^{4} }{8 K} \;, \nonumber
\eeq
where we used, as before,  $q^{2}=(q^{\prime})^{2}=\frac{3}{2}$ for our normalization of SU($2$)
charges.

Let us look a bit more closely at the factor $K$. In the asymptotically-free 
${\rm SU}(2)\! \times\! {\rm SU}(2)$ nonlinear sigma model, the mass of the FZ particles depends on the coupling $g_{0}$
as
\beq
m=\frac{C}{a}(g_{0}^{-1}e^{-2\pi/g_{0}^{2}} +\cdots) \;,\nonumber
\eeq
where $C$ is a non-universal constant, which depends on the cut-off method. Thus the second
term in $K$ according to (\ref{constant-of-field-squared}) is 
significant. Our result for the horizontal
string tension is therefore
\beq
\sigma_{\rm H} =\frac{3}{2}\left( \frac{g_{0}^{\prime}}{a}\right)^{2} \left[ 1+
\frac{4}{3}\frac{0.7296}{C^{2}\pi^{2}}\frac{(g_{0}^{\prime})^{2}}{g_{0}^{2}}e^{4\pi/g_{0}^{2}} \right]^{-1}\;.
\label{string-tension}
\eeq
Notice that for very small $g_{0}$, the exponential dominates the denominator, even if
$g_{0}^{\prime} \ll g_{0}$. Thus (\ref{string-tension}) depends on the couplings as
\beq
\sigma_{\rm H} \approx 
\frac{9C^{2}\pi^{2}}{(0.7296)8}    \, \frac{g_{0}^{2}}{a^{2}}
e^{-4\pi/g_{0}^{2}}   \;. \label{unusual}
\eeq
This expression is unusual in that all $g_{0}^{\prime}$-dependence has disappeared.

If we rewrite our expression for the string tension (\ref{string-tension}) in terms of the continuum
couplings
$e=g_{0}{\sqrt{a}}$ and $e^{\prime}=g_{0}^{\prime}{\sqrt{a}}$, we see an 
expression which is different
than (\ref{naive}). In terms of these constants and the lattice spacing,
that 
\beq
\sigma_{\rm H} =\frac{3}{2}\frac{({e}^{\prime})^{2}}{a} \left[ 1+
\frac{4}{3}\frac{0.7296}{C^{2}\pi^{2}}\frac{(e^{\prime})^{2}}{e^{2}}e^{4\pi/(e^{2}a)} \right]^{-1}
\approx 
\frac{9C^{2}\pi^{2}}{(0.7296)8}    \, \frac{e^{2}}{a}
e^{-4\pi/(e^{2}a)}  
\;,
\label{string-tension1}
\eeq
where the approximation is valid for small $a$. We 
cannot take a continuum limit of our string tension, holding $e$ and $e^{\prime}$
fixed. The reason 
the naive argument leading to (\ref{naive}) fails is in the
assumption of no ultraviolet divergences. What our method
shows is that, at least for the anisotropic range of couplings we consider, there are
such divergences. The resulting dependence
of the string tension on the couplings is not analytic. 

We believe that (\ref{string-tension}) cannot be 
extended to the isotropic regime $g_{0}^{\prime}\sim g_{0}$. To 
see why, imagine that we generalize the regularized theory to one with three couplings. In the
Euclidean Wilson lattice gauge theory, with lattice spacing $a$ and link fields 
$U_{\mu}(x)\in {\rm SU}(2)$, where $x$
lies on a three-dimensional lattice, this means an action of the form
\beq
S=\sum_{\mu\neq \nu} \frac{1}{4g_{\mu \nu}^{2}} {\rm Tr} \, U_{\mu}(x)U_{\nu}(x+{\hat \mu} a)
U_{\mu}(x+{\hat \nu}a) U_{\nu}(x) \;, \label{Wilson}
\eeq
where $\hat \mu$ denotes the unit vector in the $\mu$-direction and 
$g_{\mu \nu}=g_{\nu\mu}$. There are three distinct couplings in this model. The regime
analogous to that we consider is $g_{01}\ll g_{02}=g_{12}$. There are other regimes, we could
apply our result (\ref{string-tension}) to, namely 
$g_{02}\ll g_{01}=g_{12}$ and $g_{12}\ll g_{01}=g_{02}$. Clearly there must be a crossover
phenomenon between different behaviors of the string tension. It seems plausible that
there is other crossover behavior between these regimes to $g_{01}\sim g_{02}\sim g_{12}$
in which (\ref{naive}) takes place. Since the string tension does not depend on $g_{0}^{\prime}$
to our order of approximation, it may be that the crossover occurs at a value of 
$g_{0}^{\prime}/g_{0}$ which is not extremely small.

\section{Physical aspects of excitations}
\setcounter{equation}{0}
\renewcommand{\theequation}{7.\arabic{equation}}

In this section, we complete the picture of the confining phase by discussing the vertical
string and the nature of the pure-glue excitations. 

We already presented the basic mechanisms of linear potential between vertically-separated
quarks and the area 
decay of space-like Wilson loops in reference \cite{PhysRevD71}. We will 
show how these mechanisms fit into a general
picture of the excitations. 

Let us begin by asking what the excitations are if $g_{0}>0$ and $g_{0}^{\prime}=0$. As we
discussed in Section 3, the system splits up into decoupled layers of nonlinear sigma models. The
possible excitations are massive 
FZ particles which can travel horizontally, but not vertically. There
is, however a restriction on the these excitations, which is that the residual gauge-invariance
condition (\ref{lattice-physical})
must be satisfied. If we approximate the sum over $x^{1}$ as an integral, this condition states that
for each $x^{2}$
\beq
\left\{ \int d x^{1}\left[ j^{L}_{0}(x^{1},x^{2})_{b}-j^{R}_{0}(x^{1},x^{2}-a)_{b}\right] - g_{0}^{2}Q(x^{2})_{b} \right\}\Psi=0\;,
\label{physical}
\eeq
where $Q(x^{2})_{b}$ is the total color charge from quarks at $x^{2}$ and $\Psi$ is any physical 
state. If there are no quarks, the total right-handed charge of FZ particles in the sigma model
at $x^{2}-a$ is equal to the total left-handed charge of FZ particles in the sigma model at $x^{2}$.

Let us picture two-dimensional space partitioned into a set of parallel 
horizontal layers. Each layer contains a sigma model. The FZ particles move within
a layer, as in the left-hand side of Figure 2. Now suppose we increase $g_{0}^{\prime}$ from zero to a small value. Since
an FZ particles at $x^{2}$ has left-SU($2$) charge at $x^{2}$ and right-SU($2$) charge at
$x^{2}+a$, horizontal electric strings must join the FZ particles together. The strings
lie between the layers. Because the 
constraint (\ref{physical})
is satisfied, these strings can be consistently introduced. We now have a similar picture
of excitations, but now with strings with the tension calculated in the last section. This is shown
in the right-hand side of Figure 2. We see now that the vertical electric flux is carried by the FZ particles 
themselves - they are short segments of vertical electric flux. In fact, if we introduce
a quark and antiquark with a vertical separation, there will simply be a line of FZ particles
and horizontal strings joining them together.

The term in the effective action of the electrostatic
potential we calculated in Section 5 is 
due to charge fluctuations in the sigma model. According to the
picture we have just outlined, this means it is due to transverse (that is, vertical) fluctuations
of the string joining a quark-antiquark pair.

The reader should not be misled by the right side of Figure 2 into thinking that there is no $1$-component
of the electric field if $g_{0}^{\prime}=0$. There is electric field produced by the FZ particles in this
case, but this field carries no energy.

\begin{center}

\begin{picture}(150,60)(-10,0)

\linethickness{0.5mm}

\multiput(0,0)(0,7){8}{\multiput(0,0)(5,0){12}{\put(0,0){$-$}}}

\multiput(75,0)(0,7){8}{\multiput(0,0)(5,0){12}{\put(0,0){$-$}}}

\put(20,12){\circle{7}}
\put(10,19){\circle{7}}
\put(15,26){\circle{7}}
\put(24,33){\circle{7}}
\put(30,12){\circle{7}}
\put(40,19){\circle{7}}
\put(45,26){\circle{7}}
\put(35,33){\circle{7}}

\put(95,12){\circle{7}}
\put(85,19){\circle{7}}
\put(90,26){\circle{7}}
\put(99,33){\circle{7}}
\put(105,12){\circle{7}}
\put(115,19){\circle{7}}
\put(120,26){\circle{7}}
\put(110,33){\circle{7}}

\put(95,8.3){\line(1,0){10}}
\put(105,15.3){\line(1,0){10}}
\put(115,22.3){\line(1,0){5}}
\put(120,29.3){\line(-1,0){10}}
\put(110,36.3){\line(-1,0){11}}
\put(99,29.3){\line(-1,0){9}}
\put(90,22.3){\line(-1,0){5}}
\put(85,15.3){\line(1,0){10}}

\end{picture}

\end{center}

\vspace{5pt}


\noindent
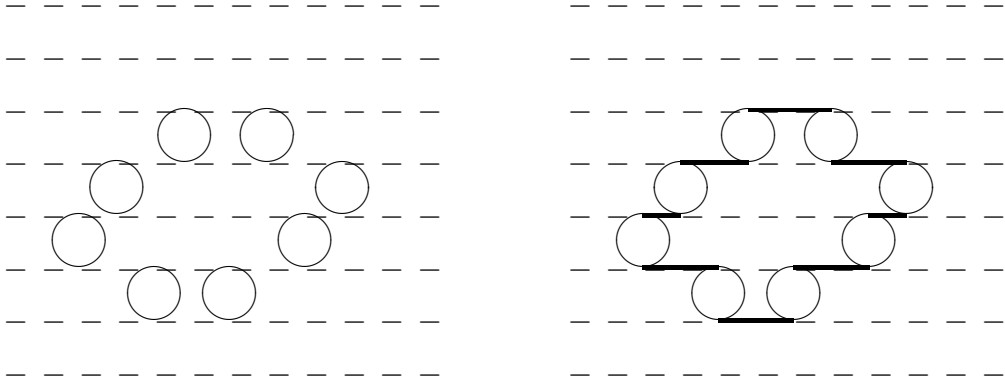
{\bf Figure 2. The circles represent FZ particles in the layers between the dashed lines. On 
the right, electric
strings connect the FZ particles. }


\vspace{10pt}

Naively, the vertical string tension is simply \cite{PhysRevD71}
\beq
\sigma_{\rm V}=\frac{m}{a}\approx \frac{C}{g_{0}a^{2}}e^{-2\pi/g_{0}^{2}} \;.\label{vertical-naive-st}
\eeq
This result follows from assuming that the energy in each layer is $m$. Strictly
speaking, the right-hand side of (\ref{vertical-naive-st})  is a lower bound to the 
vertical string tension, $\sigma_{\rm V}\ge m/a$. We expect that
there will be corrections to this 
formula. These can be found, in principle, just as we did as for the 
horizontal string 
tension. There is a subtle difference however. In the expression for the 
effective action. Instead
of (\ref{effective-action}), we must calculate
\beq
e^{{\rm i}{\mathfrak S}(\Phi)}=\lim_{T\rightarrow \infty}\left< \Psi\right\vert  {\mathcal T} 
e^{-i\int_{0}^{T} dx^{0} H_{1}} 
\left\vert \Psi \right>\;.
\label{excited-effective-action}
\eeq
The state $\Psi$ is not the vacuum. We must now have an FZ particle
in every layer between the quark and
antiquark. This problem is under study.

\section{Conclusions}

By spitting $(2+1)$-dimensional SU($2$) Yang-Mills theory anisotropically 
into integrable models, 
with an interaction between these models, we obtained an expression for the string
tension (in one direction only) at small values of the couplings. 

Two problems yet to be solved for this gauge theory are finding a
more precise result for the vertical string tension
(discussed in the last section) and the mass gap of the gauge theory. The latter problem 
is not straightforward, as the glueball excitations are collections of many FZ 
particles. Furthermore, we do not
expect the number of these FZ particles is fixed. The right approach may be to obtain a 
better understanding of the vacuum state first, then examine gauge-invariant correlators in this
state.

The effective action we used to find the horizontal string tension contains further nonlinear terms, as
we mentioned at the end of Section 5. The quartic term contains mixed correlators of left-handed
and right-handed charges of the sigma model. It is necessary to use (\ref{crossed-form-factor}), as
well as (\ref{useful-form-factor})
There is no reason why this term cannot be
determined; all that is needed is sufficient effort. Perhaps this term could shed light on 
the crossover phenomenon discussed at the end of Section 6. It will also be important to include corrections
to the mass spectrum and the form factors at this order
\cite{Delfino}.

We could extend our methods to SU($N$) gauge theories, 
if we knew the form factors for the ${\rm SU}(N)\! \times {\rm SU}(N)$
sigma model with $N>2$. Our basic idea can be formulated for any value of 
$N$ \cite{PhysRevD71}. The form-factor
problem for $N>2$ is tough, in part because 
of the presence of bound states of the fundamental
FZ particles. The S-matrix was worked out some time ago \cite{abda-wieg}, and agrees
with the Bethe {\em Ansatz} approach for an equivalent Fermionic model \cite{pol-wieg}. The 
S-matrix becomes unity in the 't$\,$Hooft limit $N\rightarrow \infty$, $g_{0}^{2}N$ fixed, and
the form factors should simplify in this limit (note: there is a large-$N$ limit of 
the ${\rm SU}(N)\! \times {\rm SU}(N)$ sigma model, whose
S-matrix is non-trivial \cite{Fateev-Kazakov-Wiegmann}. This is a
different model, because the limit is {\em not} the standard 't$\,$Hooft limit).

The picture of confinement and excitations described in Section 7 suggests that
the non-Abelian gauge theory may be dual to another field theory. The weak coupling
diagrams would correspond to the strong-coupling terms of the dual theory.

A splitting similar to the one we have used can be done in $(3+1)$ dimensions. There is
an important difference, however. As in $(2+1)$ 
dimensions, some electric-field components squared
are included
in the interaction Hamiltonian. The new feature is that the interaction Hamiltonian
also contains a magnetic-field 
component squared. Our methods would therefore not yield, strictly speaking, 
a weak-coupling result. Investigations in this direction may be of some value, nonetheless.

\section*{Acknowledgements}

I thank Poul Henrik
Damgaard and Jeff Greensite, for discussions 
about these 
ideas in their
early stages, Alexei Tsvelik, for an inspiring introductory
seminar on the applications of
form factors, and V.P. Nair, for much discourse 
on the entire project. I am grateful to Charlotte Kristjansen for the
opportunity to present these ideas at 
the Niels Bohr Summer 
Institute. I 
thank Abhishek Agarwal and Charlotte Kristjansen for comments on the manuscript. Finally I would
like to thank the referees of this paper for helpful remarks. This 
work was 
supported in part by a grant from the PSC-CUNY.

\section*{Appendix: Exact S-matrices and form factors}
\setcounter{equation}{0}
\renewcommand{\theequation}{A.\arabic{equation}}

We now present an introduction to
how S-matrices and form factors are exactly 
determined for the O($N$) nonlinear sigma model in $(1+1)$ 
dimensions. Though not a comprehensive treatment of
$(1+1)$-dimensional S-matrices and form factors, this review 
is self-contained. Since most people who might find
this paper of interest are not experts on these subjects, we 
felt it was necessary to provide complete
explanations of their less obvious aspects, which 
are not abundantly available. We 
hope this appendix is
sufficiently complete for the reader to understand the
results used in the remainder of the paper.

Integrability of the $(1+1)$-dimensional quantized
asymptotically-free O($N$) nonlinear classical 
sigma model \cite{sigma} became of interest, after it was first
established by Pohlmeyer for the classical case \cite{pohlmeyer}. Then 
general arguments were made that
the infinite set of dynamical charges can be generalized to the quantized theory 
\cite{arafyeva}. The generation of a mass gap in the $1/N$-expansion was discovered
considerably earlier \cite{stanley}.

The Lagrangian of the O($N$) sigma model depends on an $N$-component vector
field field $n=(n^{1},\dots,n^{N})$, of unit length $n\cdot n=1$:
\beq
{\mathcal L}_{NLS}=\frac{1}{2g_{0}^{2}}\eta^{\mu \nu} \partial_{\mu}n \cdot \partial_{\nu}n\;. 
\label{O(N)lagrangian}
\eeq
For $N=4$, the identification between (\ref{O(N)lagrangian}) and the 
${\rm SU}(2)\!\times\! {\rm SU}(2)$ Lagrangian (\ref{lagrangian}) is made by $U=n^{4}\ident-{\rm i}
n^{b}\sigma_{b}$.

From the $1/N$-expansion, we know that there are $N$ basic species of massive particle, which
we label by letters $j,k,l,m,\dots$ taking the values $1,2,\dots,N$. The particle states are eigenstates of 
momentum $q,p,\dots$ as well
as species $j,k,\dots$, created
on the vacuum by Faddeev-Zamolodchikov operators or FZ operators ${\mathfrak A}(q)_{j}^{\dagger}$:
\beq
\left\vert q,j,p,k,\dots \right>={\mathfrak A}(q)_{j}^{\dagger}\,{\mathfrak A}(p)_{k}^{\dagger}\cdots
\left\vert 0 \right> \;.\nonumber
\eeq
We are using the Heisenberg representation, so we work with in-states and
in-operators or out-states and out-operators.

From either the
the $1/N$-expansion or from the assumption of integrability, we 
find that particles are neither created nor destroyed 
when scattered (that is, there is no particle production). Hence multi-particle scattering may be 
decomposed as a sequence of two-particle scatterings. The requirement that this decomposition is
consistent is the Yang-Baxter equation or factorization
equation. For nonintegrable theories, there is no such decomposition.

We can find expressions for Green's functions which are valid
at large distances using exact form factors, for models with a mass gap. Green's 
functions are not exactly known, except those for the spins of the Ising
model, where {\em all} the form factors are known \cite{McCoy+co}. Nonetheless, an 
expression can be written
down for the vacuum expectation value of a product of operators which approaches the 
correct answer at large separations.

\newpage

\begin{center}
{\bf A1. The exact S-matrix of the O($N$) nonlinear sigma model}
\end{center}
\vspace{5pt}

The two-particle S-matrix should have, on fairly general intuitive grounds, the following form:
\beq
\left._{\rm out}\right< q^{\prime},j,p^{\prime},l \left\vert q,m,p,k 
\right>_{\rm in}
\!\!&\!\!=\!\!&\!\!\delta(q^{\prime}_{1}-p_{1})\delta(p^{\prime}_{1}-q_{1}) 
{\mathcal S}_{mk}^{jl}(s) \nonumber \\
\!\!&\!\!\pm\!\!&\!\! \delta(q^{\prime}_{1}-q_{1})\delta(p^{\prime}_{1}-p_{1}) {\mathcal S}_{km}^{jl}(s) \;,
\label{in-out}
\eeq
where the four-tensor in species indices ${\mathcal S}(s)$, depends on the 
center-of-mass 
energy squared $s=(q+p)^{2}$, and $\pm$ refers to Bose or Fermi statistics (we are
going to consider the former only). We introduce the other Mandelstam variable 
$t=(p^{\prime}-p)^{2}$ and the rapidities $\theta_{1}, \theta_{2}$, related to the
momenta by the standard relation $p_{0}=m\cosh \theta_{1}$, $p_{1}=m\sinh \theta_{1}$,
$q_{0}=m\cosh \theta_{2}$, $q_{1}=m\sinh \theta_{2}$. The
relative rapidity is $\theta_{12}=\theta_{2}-\theta_{1}$. 

Bound states of two particles can form only when the center-of-mass energy
is less than the total rest-mass energy. Above the the total rest-mass energy, the
energy spectrum is continuous. Thus there may be poles in $S(s)$ in the complex $s$-plane
for $s<4m^{2}$, and there is a cut in $S(s)$ for $s>4m^{2}$. Assuming crossing $s\rightleftarrows t$, there
is a cut for $t>4m^{2}$. Kinematically, $s=2m^{2}(1+\cosh\theta_{12})$ and
$t=2m^{2}(1-\cosh\theta_{12})$. Hence there are really two cuts in the complex $s$-plane;
one for $s>4m^{2}$ and another for $s<0$. A minimal analyticity assumption is that
any poles present lie on the real $s$-axis on the interval $0<s<4m^{2}$.
 
If we make the change of variable from $s$ to $\theta \equiv \theta_{12}$, we find
the analyticity structure is as follows. The $s$-plane is mapped to the region bounded
by ${\rm Im} \,\theta =0$ and ${\rm Im}\, \theta =\pi$, called the physical
strip (the physical strip is not the same as the physical region, which is the on-shell kinematically
allowed region of $\theta$). Each of the boundaries ${\rm Im}\, \theta =0,\pi$, $-\infty<{\rm Re}\,\theta<\infty$ is a 
cut. Poles lie on the ${\rm Re} \,\theta=0$
axis only. There may be poles on this axis outside the physical strip, but these do not correspond
to physical bound states (there are no bound 
states for the model we are studying). Note that crossing corresponds to
$\theta \rightarrow {\rm i}\pi-\theta$. The resolution of the identity or overcompleteness
relation in rapidity space, which is straightforwardly obtained from that in momentum space, is
\beq
\ident = \left\vert 0 \right> \! \left< 0 \right\vert
+\sum_{M=1}^{\infty} \int\frac{d\theta_{1}\cdots d\theta_{M}}{(2\pi)^{M}M!}
\left\vert \theta_{M},j_{M}\dots,\theta_{1},j_{1} \right> \! \left< \theta_{1},j_{1},\dots,\theta_{M},j_{M} 
\right\vert \;.
\label{resolution}
\eeq

The tensor ${\mathcal S}(\theta)$ can be decomposed in the
following way, following Zamolodchikov and Zamolodchikov \cite{Zamolodchikov}, \cite{zam+zam}:
\beq
{\mathcal S}_{mk}^{jl}(\theta) =\delta^{jl} \delta_{mk}S_{1}(\theta)
+\delta_{m}^{l}\delta_{k}^{j}S_{2}(\theta)
+\delta_{m}^{j}\delta_{k}^{l}S_{3}(\theta)\; .\label{decomposition1}
\eeq
We represent this decomposition pictorially as

\begin{center}
\begin{picture}(150,20)(-10,0)

\put(10,5){${\mathcal S}_{mk}^{jl}(\theta)=$}

\put(27,5){${\mathcal S}_{1}(\theta)$}

\put(33,13){$l$}
\put(33,-3){$k$}
\put(44,13.5){$j$}
\put(44,-3){$m$}

\put(35,1){\line(1,1){5}}
\put(45,1){\line(-1,1){5}}

\put(40,7){\line(1,1){5}}
\put(40,7){\line(-1,1){5}}

\put(45,5){$+\;\;{\mathcal S}_{2}(\theta)$}

\put(60,13){$l$}
\put(60,-3){$k$}
\put(71,13.5){$j$}
\put(71,-3){$m$}

\put(61.5,1.8){\line(1,1){4}}
\put(66.5,7){\line(1,1){4}}
\put(71,1.8){\line(-1,1){9}}

\put(72,5){$+\;\;{\mathcal S}_{3}(\theta)$}

\put(85,13){$l$}
\put(85,-3){$k$}
\put(96,13.5){$j$}
\put(96,-3){$m$}

\put(86,1){\line(1,1){5}}
\put(91,6){\line(-1,1){5}}
\put(97,1){\line(-1,1){5}}
\put(92,6){\line(1,1){5}}

\put(100,5){$=$}

\put(107,13){$l$}
\put(107,-3){$k$}
\put(118,13.5){$j$}
\put(118,-3){$m$}

\put(108,1){\line(1,1){10}}
\put(108,1){\vector(1,1){4}}
\put(108,1){\vector(1,1){9}}
\put(118,1){\line(-1,1){10}}
\put(118,1){\vector(-1,1){4}}
\put(118,1){\vector(-1,1){9}}


\put(106,5){$\theta$}

\end{picture}

\end{center}

\vspace{15pt}

\noindent
Each line connecting two indices in this picture is a Kronecker delta, in accordance with 
diagrammatic
lore. The rapidity $\theta$ is drawn as an angle between incoming and 
outgoing lines (though
its range is, of course, not restricted from $0$ to $\pi$). 

The functions $S_{1}(\theta)$, $S_{2}(\theta)$ and $S_{3}(\theta)$ are assumed to be
real on the real $\theta$-axis. Therefore, by the Schwartz reflection principle, 
$S_{1}(-\theta)=S_{1}(\theta)^{*}$, $S_{2}(-\theta)=S_{2}(\theta)^{*}$ and
$S_{3}(-\theta)=S_{3}(\theta)^{*}$. 

If the reader stares at the diagrammatic form for $S(\theta)$, he or she should be able
to see that under crossing, $S_{2}({\rm i}\pi-\theta)=S_{2}(\theta)$ and 
$S_{3}({\rm i}\pi-\theta)=S_{1}(\theta)$. 

There are two non-trivial conditions satisfied by the S-matrix elements which, together
with maximal analyticity, will give its complete determination. The first of these is
unitarity within the two-particle sector. This is satisfied because 
there is no particle production. Applying unitarity to (\ref{in-out}) yields
\beq
S^{gh}_{mk}(-\theta)S^{mk}_{jl}(\theta)=\delta^{g}_{j}\delta^{h}_{l}\;. \nonumber
\eeq
This expression can be worked out by multiplying Kronecker deltas together and
contracting indices, or by diagrammatic methods. In either case, we obtain the
three relations
\beq
S_{2}(\theta)S_{2}(-\theta)+S_{3}(\theta)S_{3}(-\theta)\!\!&\!\!=\!\!&\!\!1\;, \label{unitarity1} \\
S_{2}(\theta)S_{3}(-\theta)+S_{3}(\theta)S_{2}(-\theta)\!\!&\!\!=\!\!&\!\!0\;, \label{unitarity2} \\
NS_{1}(\theta)S_{1}(-\theta)+S_{1}(\theta)S_{2}(-\theta)+
S_{1}(\theta)S_{3}(-\theta)\!\!&\!\!+\!\!&\!\!S_{2}(\theta)S_{1}(-\theta) \nonumber \\
+S_{3}(\theta)S_{1}(-\theta)\!\!&\!\!=\!\!&\!\!0.\label{unitarity3} 
\eeq
The second nontrival condition is scattering factorization or the Yang-Baxter relation. This
condition makes the three- and higher-particle scattering amplitudes unique. If we imagine three classical particles scattering, there are two possible orderings in which
this scattering can occur. Each of these orderings corresponds to a certain decomposition
of three-particle S-matrix elements in terms of two-particle S-matrix elements in the quantum
theory. If either decomposition is correct, they both are; we therefore {\it identify} them. 

Let us imagine three particles scattering with particle
rapidities $\theta_{1}$, $\theta_{2}$ and
$\theta_{3}$. The relative rapidities are $\theta=\theta_{2}-\theta_{1}$, $\theta^{\prime}=\theta_{3}
-\theta_{2}$ and $\theta+\theta^{\prime}=\theta_{3}-\theta_{1}$.
The Yang-Baxter equation is shown pictorially in Figure 3. We have labeled each of the
incoming particles with a species index $i,j,k$ and each of the outgoing particles by
another species index $l,m,o$.

Now comes the real work - turning the Yang-Baxter equation into something 
manageable. Each 
side of the Yang-Baxter equation can be decomposed into twenty-seven
terms, each proportional to a six-index tensor, which is a product of three Kronecker deltas. The
left-hand and right-hand sides of the Yang-Baxter equation are decomposed pictorially in 
Figures 4 and 5, respectively. Each of the terms is equal to a product of three
functions $S_{\alpha}(\theta)$ times a product of three Kronecker deltas. We 
abbreviate products of three such functions, {\it e.g.}
$S_{2}(\theta)S_{3}(\theta+\theta^{\prime}) S_{2}(\theta^{\prime})$ as $S_{2}S_{3}S_{2}$, keeping
the arguments in the order $\theta, \theta+\theta^{\prime},\theta^{\prime}$ always. Though
there are twenty-seven terms on each side, each term must be one of 
only fifteen distinct tensors, up to the factor of three $S$'s. That is because there are exactly
fifteen ways to make a six-index tensor from a product of three Kronecker deltas.

\vspace{5pt}
\begin{center}
\begin{picture}(150,50)(0,0)

\put(13,5){$i$}
\put(24,0){$j$}
\put(24,47){$m$}
\put(13,40){$l$}
\put(56,17){$k$}
\put(56,31){$o$}

\put(25,5){\line(0,1){40}}
\put(15,10){\line(2,1){40}}
\put(15,40){\line(2,-1){40}}

\put(30,10){$\theta$}
\put(17,30){$\theta^{\prime}$}
\put(52,24){$\theta+\theta^{\prime}$}

\put(25,5){\vector(0,1){5}}
\put(25,5){\vector(0,1){35}}
\put(25,5){\vector(0,1){20}}

\put(15,10){\vector(2,1){5}}
\put(15,10){\vector(2,1){20}}
\put(15,10){\vector(2,1){35}}

\put(55,20){\vector(-2,1){5}}
\put(55,20){\vector(-2,1){20}}
\put(55,20){\vector(-2,1){35}}

\put(72,25){$=$}

\put(137,5){$k$}
\put(124,0){$j$}
\put(124,47){$m$}
\put(137,40){$o$}
\put(91,17){$i$}
\put(91,31){$l$}

\put(125,5){\line(0,1){40}}
\put(95,20){\line(2,1){40}}
\put(95,30){\line(2,-1){40}}

\put(117,10){$\theta^{\prime}$}
\put(130,30){$\theta$}
\put(87,24){$\theta+\theta^{\prime}$}

\put(125,5){\vector(0,1){5}}
\put(125,5){\vector(0,1){20}}
\put(125,5){\vector(0,1){35}}

\put(95,20){\vector(2,1){5}}
\put(95,20){\vector(2,1){20}}
\put(95,20){\vector(2,1){35}}

\put(135,10){\vector(-2,1){5}}
\put(135,10){\vector(-2,1){20}}
\put(135,10){\vector(-2,1){35}}

\end{picture}\\

\end{center}

\vspace{5pt}

\begin{center}

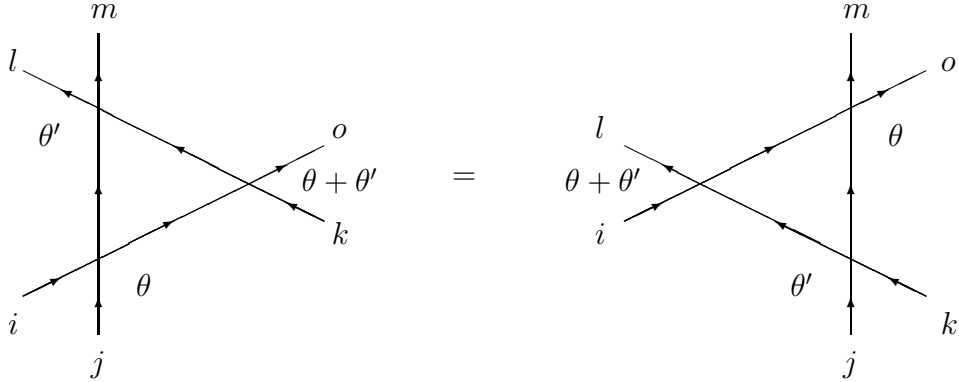
{\bf Figure 3. The factorization equation for three-particle S-matrix elements.}

\end{center}

For $N=2$, there is an additional subtlety, which is that the fifteen tensors constructed
from products of three Kronecker deltas are not linearly independent. This is important for
understanding
the sine-Gordon/massive-Thirring model, but is of no relevance to this paper. These tensors 
are linearly independent for $N\ge 3$. 

With the labeling of species as in Figure 2, we can see the form
of each term in Figures 4 and 5. For example, the first term on the left-hand side of the
Yang-Baxter 
equation is seen from Figure 4 to be $S_{1}(\theta)S_{1}(\theta+\theta^{\prime})S_{3}(\theta^{\prime})\delta_{ij}\delta^{l}_{k}\delta^{mo}$. In this way, the Yang-Baxter equation
reduces to fifteen algebraic equations. Of these fifteen equations, the seven which
are proportional to one of the following:
\beq
\delta^{l}_{i}\delta^{m}_{j}\delta^{o}_{k}\;,\;\;
\delta^{lo}\delta^{m}_{j}\delta_{ik}\;,\;\;
\delta^{l}_{k}\delta^{m}_{j}\delta^{o}_{i}\;,\;\;
\delta^{l}_{k}\delta^{mo}\delta_{ij}\;,\;\;
\delta^{lm}\delta^{o}_{i}\delta_{jk}\;,\;\;
\delta^{l}_{j}\delta^{m}_{k}\delta^{o}_{i}\;,\;\;
\delta^{l}_{k}\delta^{m}_{i}\delta^{o}_{j}\;,
\nonumber
\eeq
are
trivially satisfied. Of those eight remaining, five are redundant, leaving three non-trivial
equations \cite{Zamolodchikov},\cite{zam+zam}. The terms proportional to
$\delta^{l}_{i}\delta^{m}_{k}\delta^{o}_{j}$ and $\delta^{l}_{j}\delta^{m}_{i}\delta^{o}_{k}$
each give
\beq
S_{2}S_{3}S_{3}+S_{3}S_{3}S_{2}=S_{3}S_{2}S_{3} \;. \label{Yang-Baxter1} 
\eeq

\newpage

\vspace{5pt}
\begin{center}
\begin{picture}(150,200)(0,0)



\put(5,180){\line(0,1){20}}
\put(0,183){\line(2,1){20}}
\put(0,197){\line(2,-1){20}}

\put(25,189){$=$}

\put(34,180){\line(0,1){4}}
\put(35,186){\line(0,1){8}}
\put(36,195){\line(0,1){4}}

\put(30,182){\line(2,1){4}}
\put(35,186){\line(2,1){8}}
\put(43.25,191.5){\line(2,1){4}}

\put(31,196){\line(2,-1){4}}
\put(36,195){\line(2,-1){7}}
\put(43.25,190){\line(2,-1){4}}

\put(50,189){ ${\mathcal S}_{1}{\mathcal S}_{1}{\mathcal S}_{3}$\;\;+ }


\put(74,180){\line(0,1){4}}
\put(75,186){\line(0,1){7}}
\put(76,193.5){\line(0,1){4}}

\put(70,182){\line(2,1){4}}
\put(75,186){\line(2,1){7}}
\put(84,190.5){\line(2,1){4}}

\put(71,195){\line(2,-1){4}}
\put(76,193.5){\line(2,-1){9}}
\put(84,189.5){\line(2,-1){4}}

\put(90,189){ ${\mathcal S}_{1}{\mathcal S}_{2}{\mathcal S}_{3}$\;\;+ }


\put(114,180){\line(0,1){5}}
\put(115,186.5){\line(0,1){7}}
\put(116,194){\line(0,1){4}}

\put(110,183){\line(2,1){4}}
\put(115.25,186.75){\line(2,1){7.5}}
\put(124,190.5){\line(2,1){4}}

\put(110,196){\line(2,-1){5}}
\put(116,194){\line(2,-1){7}}
\put(124,190.5){\line(2,-1){4}}

\put(130,189){ ${\mathcal S}_{1}{\mathcal S}_{3}{\mathcal S}_{3}$}


\put(0,164){$+$}

\put(9,155){\line(0,1){4}}
\put(10,161){\line(0,1){7}}
\put(10,171){\line(0,1){4}}

\put(5,157){\line(2,1){4}}
\put(10,161){\line(2,1){7.25}}
\put(17.5,166){\line(2,1){4}}

\put(5,172){\line(2,-1){6}}
\put(11,169){\line(2,-1){6.25}}
\put(17.5,164.5){\line(2,-1){4}}

\put(25,164){ ${\mathcal S}_{1}{\mathcal S}_{1}{\mathcal S}_{2}$\;\;+ }


\put(49,155){\line(0,1){4}}
\put(50,161){\line(0,1){7}}
\put(50,171){\line(0,1){4}}

\put(45,157){\line(2,1){4}}
\put(50,161){\line(2,1){7}}
\put(60,166){\line(2,1){4}}

\put(45,172){\line(2,-1){6}}
\put(51,169){\line(2,-1){9}}
\put(60,164.5){\line(2,-1){4}}

\put(65,164){ ${\mathcal S}_{1}{\mathcal S}_{2}{\mathcal S}_{2}$\;\;+ }


\put(89,155){\line(0,1){4}}
\put(90,160){\line(0,1){7}}
\put(90,171){\line(0,1){4}}

\put(85,157){\line(2,1){4}}
\put(90,160){\line(2,1){9.5}}
\put(101,165){\line(2,1){4}}

\put(85,172){\line(2,-1){6}}
\put(91,169){\line(2,-1){8.5}}
\put(101,165){\line(2,-1){4}}

\put(105,164){ ${\mathcal S}_{1}{\mathcal S}_{3}{\mathcal S}_{2}\;\;+$}


\put(129,155){\line(0,1){4}}
\put(130,161){\line(0,1){8}}
\put(129.5,170){\line(0,1){4}}

\put(125,157){\line(2,1){4}}
\put(130,161){\line(2,1){7}}
\put(137,165.5){\line(2,1){4}}

\put(125,172){\line(2,-1){4.5}}
\put(130,169){\line(2,-1){7}}
\put(137,164.5){\line(2,-1){4}}

\put(145,164){ ${\mathcal S}_{1}{\mathcal S}_{1}{\mathcal S}_{1}$}


\put(0,139){$+$}

\put(9,130){\line(0,1){4}}
\put(10,135.5){\line(0,1){7}}
\put(10,146){\line(0,1){4}}

\put(5,132){\line(2,1){4}}
\put(10,135.5){\line(2,1){7}}
\put(20,140.5){\line(2,1){4}}

\put(5,147){\line(2,-1){7}}
\put(11,144){\line(2,-1){9}}
\put(20,139.5){\line(2,-1){4}}

\put(25,139){ ${\mathcal S}_{1}{\mathcal S}_{2}{\mathcal S}_{2}$\;\;+ }


\put(49,130){\line(0,1){4}}
\put(50.5,136){\line(0,1){7}}
\put(50,145.5){\line(0,1){4}}

\put(45,132){\line(2,1){4}}
\put(50.5,136){\line(2,1){7.5}}
\put(60,140){\line(2,1){4}}

\put(45.5,147.5){\line(2,-1){4.25}}
\put(50.5,143.5){\line(2,-1){7.5}}
\put(60,140){\line(2,-1){4}}

\put(65,139){ ${\mathcal S}_{1}{\mathcal S}_{3}{\mathcal S}_{1}$\;\;+ }


\put(90,130){\line(0,1){6}}
\put(90,136){\line(0,1){7}}
\put(91,144){\line(0,1){4}}

\put(85,133){\line(2,1){4}}
\put(91,136){\line(2,1){7}}
\put(98,140.5){\line(2,1){4}}

\put(85.5,145){\line(2,-1){4.5}}
\put(91,144){\line(2,-1){7}}
\put(98,139.5){\line(2,-1){4}}

\put(105,139){ ${\mathcal S}_{2}{\mathcal S}_{1}{\mathcal S}_{3}\;\;+$}


\put(130,130){\line(0,1){6}}
\put(130,136){\line(0,1){7}}
\put(131,144){\line(0,1){4}}

\put(125,133){\line(2,1){4}}
\put(131,136){\line(2,1){7}}
\put(140,140.5){\line(2,1){4}}

\put(125.5,145){\line(2,-1){4.5}}
\put(131,144){\line(2,-1){9}}
\put(140,139.5){\line(2,-1){4}}

\put(145,139){ ${\mathcal S}_{2}{\mathcal S}_{2}{\mathcal S}_{3}$}


\put(0,114){$+$}

\put(10,105){\line(0,1){6}}
\put(10,111){\line(0,1){7}}
\put(11,119){\line(0,1){4}}

\put(5,108){\line(2,1){4}}
\put(11,111){\line(2,1){8}}
\put(20,115){\line(2,1){4}}

\put(5.5,120){\line(2,-1){4.5}}
\put(11,119){\line(2,-1){8}}
\put(20,115){\line(2,-1){4}}

\put(25,114){ ${\mathcal S}_{2}{\mathcal S}_{3}{\mathcal S}_{3}$\;\;+ }


\put(50,105){\line(0,1){6}}
\put(50,111){\line(0,1){7}}
\put(50,121){\line(0,1){4}}

\put(45,108){\line(2,1){4}}
\put(51,111){\line(2,1){7}}
\put(58,115.5){\line(2,1){4}}

\put(45,122){\line(2,-1){6}}
\put(51,119){\line(2,-1){7}}
\put(58,114.5){\line(2,-1){4}}

\put(65,115){ ${\mathcal S}_{2}{\mathcal S}_{1}{\mathcal S}_{2}$\;\;+ }


\put(90,105){\line(0,1){6}}
\put(90,111){\line(0,1){7}}
\put(90,121){\line(0,1){4}}

\put(85,108){\line(2,1){4}}
\put(91,111){\line(2,1){7}}
\put(100,115.5){\line(2,1){4}}

\put(85,122){\line(2,-1){6}}
\put(91,119){\line(2,-1){9}}
\put(100,114.5){\line(2,-1){4}}

\put(105,115){ ${\mathcal S}_{2}{\mathcal S}_{2}{\mathcal S}_{2}\;\;+$}


\put(130,105){\line(0,1){6}}
\put(130,111){\line(0,1){7}}
\put(130,121){\line(0,1){4}}

\put(125,108){\line(2,1){4}}
\put(131,111){\line(2,1){8}}
\put(140,115){\line(2,1){4}}

\put(125,122){\line(2,-1){6}}
\put(131,119){\line(2,-1){8}}
\put(140,115){\line(2,-1){4}}

\put(145,115){ ${\mathcal S}_{2}{\mathcal S}_{3}{\mathcal S}_{2}$}


\put(0,89){$+$}

\put(10,80){\line(0,1){6}}
\put(10,86){\line(0,1){8}}
\put(9.5,95.5){\line(0,1){4}}

\put(5,83){\line(2,1){4}}
\put(10.5,86){\line(2,1){6.75}}
\put(17,90.5){\line(2,1){4}}

\put(5,98){\line(2,-1){4.5}}
\put(10,94){\line(2,-1){7}}
\put(17,89.5){\line(2,-1){4}}

\put(25,89){ ${\mathcal S}_{2}{\mathcal S}_{1}{\mathcal S}_{1}$\;\;+ }


\put(50,80){\line(0,1){6}}
\put(50,86){\line(0,1){8}}
\put(49.5,95.5){\line(0,1){4}}

\put(45,83){\line(2,1){4}}
\put(51,86){\line(2,1){6}}
\put(60,90.5){\line(2,1){4}}

\put(45,98){\line(2,-1){4.5}}
\put(50,94){\line(2,-1){14}}

\put(65,89){ ${\mathcal S}_{2}{\mathcal S}_{2}{\mathcal S}_{1}$\;\;+ }


\put(90,80){\line(0,1){6}}
\put(90,86){\line(0,1){8}}
\put(89.5,95.5){\line(0,1){4}}

\put(85,83){\line(2,1){4}}
\put(91,86){\line(2,1){7.5}}
\put(100,90){\line(2,1){4}}

\put(85,98){\line(2,-1){4.5}}
\put(90,94){\line(2,-1){8.5}}
\put(100,90){\line(2,-1){4}}

\put(105,89){ ${\mathcal S}_{2}{\mathcal S}_{3}{\mathcal S}_{1}\;\;+$}


\put(130,80.5){\line(0,1){4}}
\put(130,86){\line(0,1){7}}
\put(130.5,94){\line(0,1){4}}

\put(125,82){\line(2,1){5}}
\put(130,86){\line(2,1){6.75}}
\put(136.5,91){\line(2,1){4}}

\put(125.5,95){\line(2,-1){4.5}}
\put(130.5,94){\line(2,-1){6}}
\put(136.5,89.5){\line(2,-1){4}}

\put(145,89){ ${\mathcal S}_{3}{\mathcal S}_{1}{\mathcal S}_{3}$}


\put(0,64){$+$}

\put(11,57){\line(0,1){4}}
\put(10,61){\line(0,1){7}}
\put(10.5,69){\line(0,1){4}}

\put(5,58.75){\line(2,1){5}}
\put(11,61){\line(2,1){6}}
\put(20,65.5){\line(2,1){4}}

\put(5.5,70){\line(2,-1){4.5}}
\put(10.5,69){\line(2,-1){14}}

\put(25,64){ ${\mathcal S}_{3}{\mathcal S}_{2}{\mathcal S}_{3}$\;\;+ }


\put(51,57){\line(0,1){4}}
\put(50,61){\line(0,1){7}}
\put(50.5,69){\line(0,1){4}}

\put(45,58.75){\line(2,1){5}}
\put(51,61){\line(2,1){7.75}}
\put(60,65){\line(2,1){4}}

\put(45.5,70){\line(2,-1){4.5}}
\put(50.5,69){\line(2,-1){8.25}}
\put(60.5,65){\line(2,-1){4}}

\put(65,64){ ${\mathcal S}_{3}{\mathcal S}_{3}{\mathcal S}_{3}$\;\;+ }


\put(91,57){\line(0,1){4}}
\put(90,61){\line(0,1){7}}
\put(90,71){\line(0,1){4}}

\put(85,58.75){\line(2,1){5}}
\put(91,61){\line(2,1){7}}
\put(98,65.5){\line(2,1){4}}

\put(85,72){\line(2,-1){6}}
\put(91,69){\line(2,-1){7}}
\put(98,64.5){\line(2,-1){4}}

\put(105,64){ ${\mathcal S}_{3}{\mathcal S}_{1}{\mathcal S}_{2}\;\;+$}


\put(131,57){\line(0,1){4}}
\put(130,61){\line(0,1){7}}
\put(130,71){\line(0,1){4}}

\put(125,58.75){\line(2,1){5}}
\put(131,61){\line(2,1){7}}
\put(140,65.5){\line(2,1){4}}

\put(125,72){\line(2,-1){6}}
\put(131,69){\line(2,-1){9}}
\put(140,64.5){\line(2,-1){4}}

\put(145,64){ ${\mathcal S}_{3}{\mathcal S}_{2}{\mathcal S}_{2}$}


\put(0,39){$+$}

\put(11,32){\line(0,1){4}}
\put(10,36){\line(0,1){7}}
\put(10,46){\line(0,1){4}}

\put(5,33.75){\line(2,1){5}}
\put(11,36){\line(2,1){7.75}}
\put(20,40){\line(2,1){4}}

\put(5,47){\line(2,-1){6}}
\put(11,44){\line(2,-1){7.75}}
\put(20,40){\line(2,-1){4}}

\put(25,39){ ${\mathcal S}_{3}{\mathcal S}_{3}{\mathcal S}_{2}$\;\;+ }


\put(51,32){\line(0,1){4}}
\put(50,36){\line(0,1){8}}
\put(49.5,45.5){\line(0,1){4}}

\put(45,33.75){\line(2,1){5}}
\put(51,36){\line(2,1){6}}
\put(57,40.5){\line(2,1){4}}

\put(45,48){\line(2,-1){4.5}}
\put(50,44){\line(2,-1){7}}
\put(57,39.){\line(2,-1){4}}

\put(65,39){ ${\mathcal S}_{3}{\mathcal S}_{1}{\mathcal S}_{1}$\;\;+ }


\put(91,32){\line(0,1){4}}
\put(90,36){\line(0,1){8}}
\put(89.5,45.5){\line(0,1){4}}

\put(85,33.75){\line(2,1){5}}
\put(91,36){\line(2,1){6}}
\put(100,40.5){\line(2,1){4}}

\put(85,48){\line(2,-1){4.5}}
\put(90,44){\line(2,-1){14}}

\put(105,39){ ${\mathcal S}_{3}{\mathcal S}_{2}{\mathcal S}_{1}\;\;+$}


\put(131,32){\line(0,1){4}}
\put(130,36){\line(0,1){8}}
\put(129.5,46){\line(0,1){4}}

\put(125,33.75){\line(2,1){5}}
\put(131,36){\line(2,1){7.5}}
\put(140,40){\line(2,1){4}}

\put(125,48){\line(2,-1){4.5}}
\put(130,44){\line(2,-1){8.25}}
\put(140,40){\line(2,-1){4}}

\put(145,39){ ${\mathcal S}_{3}{\mathcal S}_{3}{\mathcal S}_{1}$}

\put(5,15){\bf Figure 4. Expansion of the left-hand side of the factorization equation.}

\end{picture}\\

\end{center}

\newpage

\vspace{5pt}
\begin{center}
\begin{picture}(150,200)(0,0)



\put(15,180){\line(0,1){20}}
\put(0,188){\line(2,1){20}}
\put(0,192){\line(2,-1){20}}

\put(25,189){$=$}

\put(43,182){\line(0,1){4}}
\put(44,186){\line(0,1){8.5}}
\put(45,196){\line(0,1){4}}

\put(32,187.5){\line(2,1){4}}
\put(36,190.5){\line(2,1){8}}
\put(45,196){\line(2,1){4}}

\put(32,192.5){\line(2,-1){4}}
\put(36,189.5){\line(2,-1){7}}
\put(44,186){\line(2,-1){4}}

\put(48,189){ ${\mathcal S}_{1}{\mathcal S}_{1}{\mathcal S}_{3}$\;\;+ }


\put(83,182){\line(0,1){4}}
\put(84,186){\line(0,1){8.5}}
\put(85,196){\line(0,1){4}}

\put(70,187.5){\line(2,1){4}}
\put(76,190.5){\line(2,1){8}}
\put(85,196){\line(2,1){4}}

\put(70,192.5){\line(2,-1){6}}
\put(76,189.5){\line(2,-1){7}}
\put(84,186){\line(2,-1){4}}

\put(88,189){ ${\mathcal S}_{1}{\mathcal S}_{2}{\mathcal S}_{3}$\;\;+ }


\put(123,182.5){\line(0,1){4}}
\put(124,186){\line(0,1){8.3}}
\put(125,196){\line(0,1){4}}

\put(109,187.5){\line(2,1){5}}
\put(116,190){\line(2,1){8}}
\put(125,196){\line(2,1){4}}

\put(109,192.5){\line(2,-1){5}}
\put(116,190){\line(2,-1){7}}
\put(124,186){\line(2,-1){4}}

\put(127,190){ ${\mathcal S}_{1}{\mathcal S}_{3}{\mathcal S}_{3}$}


\put(0,164){$+$}

\put(19,155){\line(0,1){4}}
\put(19,161){\line(0,1){8.5}}
\put(20,171){\line(0,1){4}}

\put(7,162.5){\line(2,1){4}}
\put(11,165.5){\line(2,1){8}}
\put(20,171){\line(2,1){4}}

\put(7,167.5){\line(2,-1){4}}
\put(11,164.5){\line(2,-1){9}}
\put(20,160){\line(2,-1){4}}

\put(23,164){ ${\mathcal S}_{1}{\mathcal S}_{1}{\mathcal S}_{2}$\;\;+ }


\put(59,155){\line(0,1){4}}
\put(59,161){\line(0,1){8.5}}
\put(60,171){\line(0,1){4}}

\put(45,162.5){\line(2,1){4}}
\put(51,165.5){\line(2,1){8.}}
\put(60,171){\line(2,1){4}}

\put(45,167.5){\line(2,-1){6}}
\put(51,164.5){\line(2,-1){9}}
\put(60,160){\line(2,-1){4}}

\put(63,164){ ${\mathcal S}_{1}{\mathcal S}_{2}{\mathcal S}_{2}$\;\;+ }


\put(99,155){\line(0,1){4}}
\put(99,161){\line(0,1){8.5}}
\put(100,171){\line(0,1){4}}

\put(83,162.5){\line(2,1){5}}
\put(90,165){\line(2,1){8.6}}
\put(100,171){\line(2,1){4}}

\put(83,167.5){\line(2,-1){5}}
\put(90,165){\line(2,-1){11}}
\put(100,160){\line(2,-1){4}}

\put(103,164){ ${\mathcal S}_{1}{\mathcal S}_{3}{\mathcal S}_{2}\;\;+$}


\put(140,155){\line(0,1){4}}
\put(139,160.5){\line(0,1){9}}
\put(140,171){\line(0,1){4}}

\put(127,162.5){\line(2,1){4}}
\put(131,165.5){\line(2,1){8.2}}
\put(140,171){\line(2,1){4}}

\put(127,167.5){\line(2,-1){4}}
\put(131,164.5){\line(2,-1){8}}
\put(140,159){\line(2,-1){4}}

\put(143,164){ ${\mathcal S}_{1}{\mathcal S}_{1}{\mathcal S}_{1}$}


\put(0,139){$+$}

\put(19,130){\line(0,1){4}}
\put(19,136){\line(0,1){8.5}}
\put(20,146){\line(0,1){4}}

\put(5,137.5){\line(2,1){4}}
\put(11,140.5){\line(2,1){8.25}}
\put(20,146){\line(2,1){4}}

\put(5,142.5){\line(2,-1){6}}
\put(11,139.5){\line(2,-1){11}}
\put(20,135){\line(2,-1){4}}

\put(23,139){ ${\mathcal S}_{1}{\mathcal S}_{2}{\mathcal S}_{2}$\;\;+ }


\put(60,130){\line(0,1){4}}
\put(59,135.75){\line(0,1){8.5}}
\put(60,146){\line(0,1){4}}

\put(43,137.5){\line(2,1){5}}
\put(50,140){\line(2,1){9}}
\put(60,146){\line(2,1){4}}

\put(43,142.5){\line(2,-1){5}}
\put(50,140){\line(2,-1){9}}
\put(60,134){\line(2,-1){4}}

\put(63,139){ ${\mathcal S}_{1}{\mathcal S}_{3}{\mathcal S}_{1}$\;\;+ }


\put(98,132){\line(0,1){4}}
\put(99,136){\line(0,1){11}}
\put(99,145){\line(0,1){4}}

\put(87,137.5){\line(2,1){4}}
\put(91,140.5){\line(2,1){7}}
\put(100,145){\line(2,1){4}}

\put(87,142.5){\line(2,-1){4}}
\put(91,139.5){\line(2,-1){7}}
\put(99,136){\line(2,-1){4}}

\put(103,139){ ${\mathcal S}_{2}{\mathcal S}_{1}{\mathcal S}_{3}\;\;+$}


\put(138,132){\line(0,1){4}}
\put(139,136){\line(0,1){11}}
\put(139,146){\line(0,1){4}}

\put(125,137.5){\line(2,1){4}}
\put(131,140.5){\line(2,1){7}}
\put(140,145){\line(2,1){4}}

\put(125,142.5){\line(2,-1){6}}
\put(131,139.5){\line(2,-1){7}}
\put(139,136){\line(2,-1){4}}

\put(143,139){ ${\mathcal S}_{2}{\mathcal S}_{2}{\mathcal S}_{3}$}


\put(0,114){$+$}

\put(18,107){\line(0,1){4}}
\put(19,111){\line(0,1){11}}
\put(19,121){\line(0,1){4}}

\put(3,112.5){\line(2,1){5}}
\put(10,115){\line(2,1){7}}
\put(20,120){\line(2,1){4}}

\put(3,117.5){\line(2,-1){5}}
\put(10,115){\line(2,-1){8}}
\put(19,111){\line(2,-1){4}}

\put(23,114){ ${\mathcal S}_{2}{\mathcal S}_{3}{\mathcal S}_{3}$\;\;+ }


\put(59,105){\line(0,1){4}}
\put(59,111){\line(0,1){11}}
\put(59,121){\line(0,1){4}}

\put(47,112.5){\line(2,1){4}}
\put(51,115.5){\line(2,1){7}}
\put(60,120){\line(2,1){4}}

\put(47,117.5){\line(2,-1){4}}
\put(51,114.5){\line(2,-1){9}}
\put(60,110){\line(2,-1){4}}

\put(63,114){ ${\mathcal S}_{2}{\mathcal S}_{1}{\mathcal S}_{2}$\;\;+ }


\put(99,105){\line(0,1){4}}
\put(99,111){\line(0,1){11}}
\put(99,121){\line(0,1){4}}

\put(85,112.5){\line(2,1){4}}
\put(91,115.5){\line(2,1){7}}
\put(100,120){\line(2,1){4}}

\put(85,117.5){\line(2,-1){6}}
\put(91,114.5){\line(2,-1){9}}
\put(100,110){\line(2,-1){4}}

\put(103,114){ ${\mathcal S}_{2}{\mathcal S}_{2}{\mathcal S}_{2}\;\;+$}


\put(139,105){\line(0,1){4}}
\put(139,111){\line(0,1){11}}
\put(139,121){\line(0,1){4}}

\put(123,112.5){\line(2,1){5}}
\put(130,115){\line(2,1){7}}
\put(140,120){\line(2,1){4}}

\put(123,117.5){\line(2,-1){5}}
\put(130,115){\line(2,-1){11}}
\put(140,110){\line(2,-1){4}}

\put(143,115){ ${\mathcal S}_{2}{\mathcal S}_{3}{\mathcal S}_{2}$}


\put(0,89){$+$}

\put(20,80){\line(0,1){4}}
\put(19,85.5){\line(0,1){11}}
\put(19,96){\line(0,1){4}}

\put(7,87.5){\line(2,1){4}}
\put(11,90.5){\line(2,1){7}}
\put(20,94.5){\line(2,1){4}}

\put(7,92.5){\line(2,-1){4}}
\put(11,89.5){\line(2,-1){8}}
\put(20,84){\line(2,-1){4}}

\put(23,89){ ${\mathcal S}_{2}{\mathcal S}_{1}{\mathcal S}_{1}$\;\;+ }


\put(60,80){\line(0,1){4}}
\put(59,85.5){\line(0,1){11}}
\put(59,96){\line(0,1){4}}

\put(45,87.5){\line(2,1){4}}
\put(51,90.5){\line(2,1){7}}
\put(60,94.5){\line(2,1){4}}

\put(45,92.5){\line(2,-1){6}}
\put(51,89.5){\line(2,-1){8}}
\put(60,84){\line(2,-1){4}}

\put(63,89){ ${\mathcal S}_{2}{\mathcal S}_{2}{\mathcal S}_{1}$\;\;+ }


\put(100,80){\line(0,1){4}}
\put(99,85.5){\line(0,1){11}}
\put(99,96){\line(0,1){4}}

\put(83,87.5){\line(2,1){5}}
\put(90,90){\line(2,1){7}}
\put(100,94.5){\line(2,1){4}}

\put(83,92.5){\line(2,-1){5}}
\put(90,90){\line(2,-1){9}}
\put(100,84){\line(2,-1){4}}

\put(103,89){ ${\mathcal S}_{2}{\mathcal S}_{3}{\mathcal S}_{1}\;\;+$}


\put(138,82){\line(0,1){4}}
\put(139,86){\line(0,1){7}}
\put(138,93.5){\line(0,1){4}}

\put(127,87.5){\line(2,1){4}}
\put(131,90.){\line(2,1){7}}
\put(139,93.){\line(2,1){4}}

\put(127,92.5){\line(2,-1){4}}
\put(131,89.5){\line(2,-1){7}}
\put(139,86){\line(2,-1){4}}

\put(143,89){ ${\mathcal S}_{3}{\mathcal S}_{1}{\mathcal S}_{3}$}


\put(0,64){$+$}

\put(18,57){\line(0,1){4}}
\put(19,61){\line(0,1){7}}
\put(18,69){\line(0,1){4}}

\put(5,62.5){\line(2,1){4}}
\put(11,65.5){\line(2,1){7}}
\put(19,68){\line(2,1){4}}

\put(5,67.5){\line(2,-1){6}}
\put(11,64.5){\line(2,-1){7}}
\put(19,61){\line(2,-1){4}}

\put(23,64){ ${\mathcal S}_{3}{\mathcal S}_{2}{\mathcal S}_{3}$\;\;+ }


\put(57,57.5){\line(0,1){4}}
\put(59,61){\line(0,1){7}}
\put(58,69){\line(0,1){4}}

\put(43,62.5){\line(2,1){5}}
\put(50,65){\line(2,1){8}}
\put(59,68){\line(2,1){4}}

\put(43,67.5){\line(2,-1){5}}
\put(50,65){\line(2,-1){7}}
\put(59,61){\line(2,-1){4}}

\put(63,64){ ${\mathcal S}_{3}{\mathcal S}_{3}{\mathcal S}_{3}$\;\;+ }


\put(99,55){\line(0,1){4}}
\put(99,61){\line(0,1){7}}
\put(98,69){\line(0,1){4}}

\put(87,62.5){\line(2,1){4}}
\put(91,65.5){\line(2,1){7}}
\put(99,68){\line(2,1){4}}

\put(87,67.5){\line(2,-1){4}}
\put(91,64.5){\line(2,-1){11}}
\put(100,60){\line(2,-1){4}}

\put(103,64){ ${\mathcal S}_{3}{\mathcal S}_{1}{\mathcal S}_{2}\;\;+$}


\put(139,55){\line(0,1){4}}
\put(139,61){\line(0,1){7}}
\put(138,69){\line(0,1){4}}

\put(125,62.5){\line(2,1){4}}
\put(131,65.5){\line(2,1){7}}
\put(139,68){\line(2,1){4}}

\put(125,67.5){\line(2,-1){6}}
\put(131,64.5){\line(2,-1){11}}
\put(140,60){\line(2,-1){4}}

\put(143,64){ ${\mathcal S}_{3}{\mathcal S}_{2}{\mathcal S}_{2}$}


\put(0,39){$+$}

\put(19,30){\line(0,1){4}}
\put(19,36){\line(0,1){7}}
\put(18,44){\line(0,1){4}}

\put(3,37.5){\line(2,1){5}}
\put(10,40){\line(2,1){8}}
\put(19,43){\line(2,1){4}}

\put(3,42.5){\line(2,-1){5}}
\put(10,40){\line(2,-1){11}}
\put(20,35){\line(2,-1){4}}

\put(23,39){ ${\mathcal S}_{3}{\mathcal S}_{3}{\mathcal S}_{2}$\;\;+ }


\put(60,30){\line(0,1){4}}
\put(59,35.5){\line(0,1){7.5}}
\put(58,44){\line(0,1){4}}

\put(47,37.5){\line(2,1){4}}
\put(51,40.5){\line(2,1){7}}
\put(59,43){\line(2,1){4}}

\put(47,42.5){\line(2,-1){4}}
\put(51,39.5){\line(2,-1){8}}
\put(60,34){\line(2,-1){4}}

\put(63,39){ ${\mathcal S}_{3}{\mathcal S}_{1}{\mathcal S}_{1}$\;\;+ }


\put(100,30){\line(0,1){4}}
\put(99,35.5){\line(0,1){7.5}}
\put(98,44){\line(0,1){4}}

\put(85,37.5){\line(2,1){4}}
\put(91,40.5){\line(2,1){7}}
\put(99,43){\line(2,1){4}}

\put(85,42.5){\line(2,-1){6}}
\put(91,39.5){\line(2,-1){8}}
\put(100,34){\line(2,-1){4}}

\put(103,39){ ${\mathcal S}_{3}{\mathcal S}_{2}{\mathcal S}_{1}\;\;+$}


\put(140,30){\line(0,1){4}}
\put(139,35.5){\line(0,1){7.5}}
\put(138,44){\line(0,1){4}}

\put(123,37.5){\line(2,1){5}}
\put(130,40){\line(2,1){8}}
\put(139,43){\line(2,1){4}}

\put(123,42.5){\line(2,-1){5}}
\put(130,40){\line(2,-1){9}}
\put(140,34){\line(2,-1){4}}

\put(143,39){ ${\mathcal S}_{3}{\mathcal S}_{3}{\mathcal S}_{1}$}

\put(5,15){\bf Figure 5. Expansion of the right-hand side of the factorization equation.}

\end{picture}\\

\end{center}

\newpage

\noindent
The terms proportional to $\delta^{lo}\delta^{m}_{i}\delta_{jk}$ and
$\delta^{lm}\delta_{ik}\delta^{o}_{j}$ each give
\beq
S_{2}S_{1}S_{1}+S_{3}S_{2}S_{1}=S_{3}S_{1}S_{2} \;. \label{Yang-Baxter2} 
\eeq
The terms proportional to $\delta^{l}_{j}\delta^{mo}\delta^{ik}$ and 
$\delta^{lo}\delta^{m}_{k}\delta_{ij}$, 
also give (\ref{Yang-Baxter2}), but with 
the arguments $\theta$ and $\theta^{\prime}$ reversed. Finally, the terms proportional
to $\delta^{l}_{i}\delta^{mo}\delta_{jk}$ and $\delta^{lo}\delta^{n}_{k}\delta^{ij}$
each give
\beq
\!\!\!NS_{1}S_{3}S_{1}+S_{1}S_{3}S_{2}+S_{1}S_{3}S_{3} 
+S_{1}S_{2}S_{1}+S_{2}S_{3}S_{1}+S_{3}S_{3}S_{1}
+S_{1}S_{1}S_{1}\!=\!S_{3}S_{1}S_{3} , \label{Yang-Baxter3} 
\eeq
In each of (\ref{Yang-Baxter1}), (\ref{Yang-Baxter2}) and (\ref{Yang-Baxter3})
the arguments are $\theta$, $\theta+\theta^{\prime}$ and $\theta^{\prime}$, respectively.

Next, let us solve the unitarity and factorization equations. If the function $h(\theta)$ is
defined as $h(\theta)=S_{2}(\theta)/S_{3}(\theta)$, (\ref{unitarity1}) becomes
\beq
h(\theta)+h(\theta^{\prime})=h(\theta+\theta^{\prime})\;.\nonumber
\eeq
Unless $h(\theta)$ vanishes, the only possible solution is 
\beq
h(\theta)=\frac{{\rm i}}{\lambda}\theta\;,  \label{ratio}
\eeq
for
some constant $\lambda$. Note that (\ref{unitarity2}) yields $h(-\theta)+h(\theta)=0$, which is
automatically satisfied. Defining another function $\rho(\theta)=S_{1}(\theta)/S_{3}(\theta)$, 
equation (\ref{Yang-Baxter2}) and (\ref{ratio}) imply that
\beq
\rho(\theta+\theta^{\prime})^{-1}+\rho(\theta^{\prime})^{-1}=-\frac{{\rm i}\theta}{\lambda}\;.
\nonumber
\eeq
The solution for $\rho(\theta)$ is 
\beq
\rho(\theta)=-\frac{{\rm i}\lambda}{{\rm i}\kappa-\theta} \;, \label{ratio1}
\eeq
where $\kappa$ is another constant. In terms of the original functions,
\beq
S_{3}(\theta)=\frac{-{\rm i}\lambda}{\theta}S_{2}(\theta)\;,\;\;
S_{1}(\theta)=-\frac{{\rm i}\lambda}{{\rm i}\kappa-\theta} S_{2}(\theta)\;. \label{S_1-and-S_3}
\eeq
Next we substitute (\ref{S_1-and-S_3}) into (\ref{Yang-Baxter3}) and multiply both
sides by
$\theta({\rm i}\kappa-\theta)(\theta+\theta^{\prime})({\rm i}\kappa-\theta-\theta^{\prime})\theta^{\prime}
({\rm i}\kappa-\theta^{\prime})$. The result is that a fourth-order polynomial in rapidities $\theta$
and $\theta^{\prime}$ is zero. The zeroth-, first- and fourth-order terms are identically zero. Both
the second- and third-order terms are zero provided
\beq
\kappa=\frac{\lambda(N-2)}{2} \;.\nonumber
\eeq
By the crossing property and (\ref{S_1-and-S_3}), we have that $\kappa=\pi$ and thus
$\lambda=2\pi/(N-2)$. By (\ref{unitarity1}) and (\ref{S_1-and-S_3}), we obtain
\beq
S_{2}(\theta)S_{2}(-\theta)\!\!&\!\!=\!\!&\!\!\frac{\theta^{2}}{\theta^{2}+\frac{4\pi^{2}}{(N-2)^{2}}}\;, 
\nonumber \\
S_{3}(\theta)=-\frac{2\pi{\rm i}}{(N-2)\theta}S_{2}(\theta)\;\!\!&\!\!,\!\!&\!\!\;\;
S_{1}(\theta)=-\frac{{2\pi\rm i}}{(N-2)({\rm i}\pi-\theta)} S_{2}(\theta)\;. \label{near-solution}
\eeq
Using crossing, we can write the first of these equations as
\beq
S_{2}(\theta)S_{2}(\pi{\rm i}+\theta)=\frac{\theta^{2}}{\theta^{2}+\frac{4\pi^{2}}{(N-2)^{2}}}\;.
\label{near-solution-prime}
\eeq

We will solve (\ref{near-solution-prime}) for the solution of the two-particle S-matrix inside the
physical strip, assuming maximal analyticity. Rather than following references
\cite{Zamolodchikov} and \cite{zam+zam} at this stage, we will instead use an elegant
prescription invented by Karowski, Thun, Truong and Weisz for the sine-Gordon/massive-Thirring 
model \cite{KTTW}. We use this prescription not only because it is straightforward, but because it
gives the solution in precisely the form we need for obtaining form 
factors of current operators. We 
are not aware of this method being used for models other than 
sine-Gordon
in the literature (we suspect it lies buried in the notes 
of the serious practitioners of the subject), but 
the 
result is certainly well known, and 
can be obtained by other methods.

To use the prescription of reference \cite{KTTW}, it is convenient to decompose 
$S(\theta)$ differently. Instead of (\ref{decomposition1}), we write
\beq
S(\theta)=P_{0}S_{0}(\theta)+P_{S}S_{S}(\theta)+P_{A}S_{A}(\theta)\;, \label{decomposition2}
\eeq
where $P_{0}$, $P_{S}$ and $P_{A}$ project to the singlet, symmetric-traceless and antisymmetric
irreducible representations, respectively, on $N\otimes N$ :
\beq
(P_{0})^{jl}_{mk}\!\!&\!\!=\!\!&\!\!\frac{1}{N}\delta^{jl}\delta_{mk}\;,\;\;
(P_{S})^{jl}_{mk}=\frac{1}{2}\left( \delta^{j}_{k}\delta^{l}_{m}-\delta^{j}_{m}\delta^{l}_{k}\right)\;, 
\nonumber\\
(P_{A})^{jl}_{mk}\!\!&\!\!=\!\!&\!\!\frac{1}{2}\left( \delta^{j}_{k}\delta^{l}_{m}+\delta^{j}_{m}\delta^{l}_{k}\right)
-\frac{1}{N}\delta^{jl}\delta_{mk}\;.
\label{representation-projections}
\eeq
The equations (\ref{near-solution}), (\ref{near-solution-prime}) become
\beq
S_{A}(\theta)S_{A}(\pi{\rm i}+\theta)\!\!&\!\!=\!\!&\!\!\left(1+\frac{{\rm i}\lambda}{\theta}\right)
\left(1+\frac{{\rm i}\lambda}{\theta+\pi{\rm i}}\right)
\frac{\theta^{2}}{\theta^{2}+\lambda^{2}}=
\frac{1+\frac{{\rm i}\lambda}{\theta+{\rm i}\pi}}{1-\frac{{\rm i}\lambda}{\theta}}\;, 
\nonumber 
\eeq
\beq
S_{0}(\theta)=\!\!
\frac{(\theta-{\rm i}\lambda)(\theta-{\rm i}\pi)+
{\rm i}N\lambda\theta}{(\theta+{\rm i}\lambda)
(\theta-{\rm i}\pi)}S_{A}(\theta)
\;,\;\;
S_{S}(\theta)=\frac{\theta-{\rm i}\lambda}{\theta+{\rm i}\lambda} S_{A}(\theta)\;, \label{near-solution1}
\eeq
where, as before, $\lambda=2\pi/(N-2)$. We will use the first of (\ref{near-solution1}) to solve for
$S_{A}(\theta)$. 

Following reference \cite{KTTW} we assume \begin{enumerate}
\item The function $S_{A}(\theta)$ is analytic\footnote{In the case of the sine-Gordon model, there
are bound-state poles on the imaginary
axis in the analogous S-matrix element, 
for some choices of the coupling 
\cite{zamS-G}, \cite{KarowskiThun}, \cite{zam+zam}, \cite{BergReview}. Hence this 
analyticity
assumption is only valid for the case where the solitons have
repulsive interaction. After the answer is found for this case, it can be generally applied by
analytic continuation \cite{KTTW}.} and non-zero in the interior of the physical strip
(it has, of course, cuts on the boundaries). 
\item $\vert \ln S_{A}(\theta)/\sinh (z-\theta) \vert \rightarrow 0$
as $\vert {\rm Re}\,z \vert \rightarrow \infty$, for any fixed choice of $\theta$ in the
physical strip.
\end{enumerate}

Now $\sinh(z-\theta)$ has only one zero of $z$ in the physical strip, provided $\theta$ is in the 
physical strip. Let $C$ be the counter-clockwise contour enclosing the physical strip; so C extends from $-\infty$ to $\infty$ and from
$\pi{\rm i}+\infty$ to $\pi{\rm i}-\infty$. Then
\beq
\ln\, S_{A}(\theta)\!\!&\!\!=\!\!&\!\!\frac{1}{2\pi{\rm i}}\int_{C}\frac{dz}{\sinh(z-\theta)} \,\ln S_{A}(z) 
\nonumber \\
\!\!&\!\!=\!\!&\!\! \frac{1}{2\pi{\rm i}}\int_{-\infty}^{\infty}\frac{dz}{\sinh(z-\theta)}
\,\ln [S_{A}(z)S_{A}(z+\pi{\rm i})] \nonumber \\
\!\!&\!\!=\!\!&\!\! \frac{1}{2\pi{\rm i}}\int_{-\infty}^{\infty}\frac{dz}{\sinh(z-\theta)}
\,\ln \frac{1+\frac{{\rm i}\lambda}{z+{\rm i}\pi}}{1-\frac{{\rm i}\lambda}{z}}
\label{almost-the-answer}
\eeq
where in the last step, we used  (\ref{near-solution1}). 

Though (\ref{almost-the-answer}) is a succinct expression for the S-matrix, it is not yet
the most useful form. To find a better form, we need the following Fourier transforms
\beq
\int_{-\infty}^{\infty}\frac{dz}{2\pi}\,\frac{e^{{\rm i}\xi z}}{\sinh(z-\theta)}
=\frac{{\rm i}e^{{\rm i}(\theta-\pi{\rm i}/2)}}{2\cosh\frac{\pi \xi}{2}}\;,\;\;
\int_{-\infty}^{\infty}\frac{dz}{2\pi}\,e^{{\rm i}\xi z} \,\ln \left(1-\frac{{\rm i}\lambda}{z}\right)
=\frac{1}{\xi}\left( 1-e^{-\xi \lambda}\right) \;,\nonumber
\eeq
which can be worked out using basic complex-integration methods. Substituting these Fourier
transforms into (\ref{almost-the-answer}), we obtain
\beq
S_{A}(\theta)=\exp \,\int_{0}^{\infty}\frac{d\xi}{\xi}\frac{e^{-\frac{2\xi}{N-2} } -1}{e^{\xi}+1}
\sinh\left( \frac{\xi\theta}{\pi{\rm i}} \right)\;. \label{the-S-matrix!}
\eeq
This is the form of the S-matrix we shall use to discuss form factors. We note that this
expression (\ref{the-S-matrix!}) can be converted to the Zamolodchikovs' rational expression
of gamma functions \cite{Zamolodchikov}, \cite{zam+zam} using the integral formula \cite{W+W}
\beq
\Gamma(z)=\exp \int_{0}^{\infty} \frac{d\xi}{\xi} \left[
\frac{e^{-\xi z }-e^{-\xi} }{1-e^{-\xi}}+(z-1)e^{-\xi}
\right]\;, \;\; {\rm Re}\,z>0\;,  \nonumber
\eeq
which can be checked by differentiation and comparison with the integral formula for the
psi function and verification
that the right-hand side is equal to one at $z=1$. This integral formula was first used by 
Weisz \cite{Weisz}
to write the infinite-gamma-function-product-ratio  expressions of
Zamolodchikov \cite{zamS-G}, \cite{zam+zam}
for the sine-Gordon S-matrix in a more compact form. 

We have obtained the {\it minimal} O($N$)-symmetric S-matrix, {\it i.e.} that with as much
analyticity as possible. When expanded in powers of
$1/N$, this agrees with the S-matrix of the nonlinear sigma model obtained by standard
$1/N$-expansion methods
\cite{Zamolodchikov}, \cite{1/N}. If the two-particle S-matrix elements are multiplied
by CDD factors \cite{CDD}
\beq
\prod_{k}\frac{\sinh\theta+{\rm i}\sin\alpha_{k}}{\sinh\theta-{\rm i}\sin\alpha_{k}}\;, \nonumber
\eeq
unitarity and factorization are unaffected. One
such S-matrix obtained in this way is that of the O($N$) Gross-Neveu model. See references
\cite{Zamolodchikov}, \cite{zam+zam} for more discussion. The supersymmetric sigma model
is a nonlinear sigma model and a Gross-Neveu model coupled together; its
S-matrix was obtained in reference \cite{ShankarWitten}.

\vspace{5pt}
\begin{center}
{\bf A2. Exact form factors}
\end{center}
\vspace{5pt}

The determination of exact two-particle form factors was initiated by Vergeles and
Gryanik \cite{VergelesGryanik} for the fundamental particles of the
sinh-Gordon model and by Weisz \cite{Weisz} for the
solitons of
the sine-Gordon model. Subsequently, Karowski and Wiesz \cite{KarowskiWeisz} obtained generalizations of this
result for the sine-Gordon model and extended the method to other models. Smirnov and later
Kirillov and Smirnov
\cite{Smirnov} found extensions to higher-point form factors. Smirnov 
formulated
a set of axioms underlying the entire subject \cite{SmirnovBook} and which
was useful in studying specific field theories. In the meantime, 
Smirnov's axioms have actually been proved as theorems, assuming maximal analyticity
and the LSZ axioms \cite{BabKarow}. Though establishing their validity from deeper
principles is certainly
worthwhile, the validity
of Smirnov's axioms can be argued from symmetries, crossing, integrability
and a little physical intuition.


Our interest in form factors is that they contain off-shell information about integrable quantum
field theories. This means that they can be used to study deformations of such
theories which are no longer integrable. The form-factor program has lead to nonperturbative
calculations 
in statistical
mechanics \cite{McCoy+co}, \cite{FonsecaZamolodchikov} and condensed-matter physics
\cite{BhaseenTsvelick}, \cite{EsslerKonik}
which agree well with experimental measurements.

A form factor is a matrix element of an operator ${\mathfrak B}(x)$ between multi-particle states. We
can obtain all of these from
\beq
f^{\mathfrak B}(\theta_{1},\dots,\theta_{M})_{j_{1}\cdots j_{M}}
=\left<0\right\vert {\mathfrak B}(0) \left\vert \theta_{M},j_{M},\dots,\theta_{1},j_{1}\right>_{\rm in}
\label{form-factor-def}
\eeq
by crossing and the Lorentz-transformation properties of the operator ${\mathfrak B}(x)$. For 
example
\beq
f^{\mathfrak B}(\theta_{1},\dots,\theta_{M})_{j_{1}\cdots j_{M}} \exp -{\rm i}\sum_{l=1}^{M}p_{j_{l}}\cdot
x_{l}
=\left<0\right\vert {\mathfrak B}(x) \left\vert \theta_{M},j_{M},\dots,\theta_{1},j_{1}\right>_{\rm in}
\label{form-factor-x}
\eeq

Let us state some basic commutation relations of the FZ particle
creation in-operators and annihilation
in-operators
${\mathfrak A}_{j}(\theta)^{\dagger}$ and ${\mathfrak A}_{j}(\theta)$, respectively
(the arguments of these operators are now rapidities instead of momenta):
\beq
{\mathfrak A}_{i}(\theta_{2})^{\dagger}{\mathfrak A}_{j}(\theta_{1})^{\dagger}
\!\!&\!\!=\!\!&\!\!S^{lm}_{ij}(\theta_{2}-\theta_{1}){\mathfrak A}_{m}(\theta_{1})^{\dagger}
{\mathfrak A}_{l}(\theta_{2})^{\dagger}\;, \nonumber
\\
{\mathfrak A}_{i}(\theta_{2}){\mathfrak A}_{j}(\theta_{1})
&\!\!=\!\!&\!\!S^{lm}_{ij}(\theta_{2}-\theta_{1}){\mathfrak A}_{m}(\theta_{1}){\mathfrak A}_{l}(\theta_{2})\;, \nonumber
\\
{\mathfrak A}_{i}(\theta_{2}){\mathfrak A}_{j}(\theta_{1})^{\dagger}
\!\!&\!\!=\!\!&\!\!2\pi\delta_{ij}\delta(\theta_{2}-\theta_{1})+
S^{mi}_{jl}(\theta_{2}-\theta_{1})
{\mathfrak A}_{m}(\theta_{1})^{\dagger}{\mathfrak A}_{l}(\theta_{2})\;. \label{FZequations}
\eeq
The third equation is obeyed for a free
field theory with the S-matrix element replaced by the identity on 
two-particle space. The $2\pi$ in the normalization of the first term on the right-hand
side of this equation follows from the 
overcompleteness
relation (\ref{resolution}) in rapidity space. The 
form of the second term on the right-hand side of this equation follows from crossing. 

For pedagogical reasons, we shall 
list the form-factor axioms for general multi-particle form 
factors and attempt to
convince the reader that
they are reasonable. Our list follows Essler and 
Konik \cite{EsslerKonik}, though we attempt to provide 
more justification
for some of the axioms. We use only 
four of these five axioms. We
will only consider the case of operators with 
mutually-local commutation relations. For discussion
of how to deal with other commutation relations, see 
Chapter 6 of Smirnov's book \cite{SmirnovBook}
or the article by Essler and Konik \cite{EsslerKonik}. 

\vspace{5pt}

\noindent
1. {\em Scattering Axiom}. This follows from the properties (\ref{FZequations}) of the FZ operators. It
is 
\beq
&\!\!f\!\!&\!\!^{\mathfrak B}(\theta_{1},\dots,\theta_{i-1},\theta_{i+1},\theta_{i},
\theta_{i+2}\dots, \theta_{M})_{j_{1}\cdots j_{i-1} j_{i+1} j_{i} j_{i+2}\cdots j_{M}}    \nonumber \\
&\!\!=\!\!& 
S^{k_{i} k_{i+1}}_{j_{i} j_{i+1}}(\theta_{i}-\theta_{i+1})
f^{\mathfrak B}(\theta_{1},\dots,\theta_{i-1},\theta_{i},
\theta_{i+1},
\theta_{i+2},\dots, \theta_{M})_{j_{1}\cdots j_{i-1} k_{i} k_{i+1} j_{i+2}\cdots j_{M}}  .   
\;\;\;\;\;\;\;\;\;\;\;
\label{axiom1}
\eeq
This axiom is sometimes called Watson's theorem \cite{KWatson}. 

\vspace{5pt}

\noindent
2. {\em Periodicity Axiom}. This axiom is subtle. It 
is motivated by crossing. The axiom states
\beq
f^{\mathfrak B}(\theta_{1},\dots,\theta_{M})_{j_{1}\cdots j_{M}}
=f^{\mathfrak B}(\theta_{M}-2\pi {\rm i},\theta_{1}\dots,\theta_{M-1})_{j_{M} j_{1}\cdots j_{M-1}}\;.
\label{axiom2}
\eeq
To see where this axiom comes from, let us see what happens when a creation FZ operator
in front of a ket (state vector) is replaced by an annihilation operator behind a bra (dual state
vector) by crossing. Consider the Green's function of FZ operators and $\mathfrak B$
\beq
\left< \!\!\! \right. \!\!\!&0\!\!\!\!&\!\! \left. \right\vert {\mathfrak A}_{j_{1}}(\theta_{1})\; {\mathfrak B}(0)\;
{\mathfrak A}_{j_{M}}(\theta_{M})^{\dagger}{\mathfrak A}_{j_{M-1}}(\theta_{M-1})^{\dagger}
\cdots
{\mathfrak A}_{j_{2}}(\theta_{2})^{\dagger} \left\vert 0\right>_{\rm connected}  \nonumber \\ 
&\!\!=\!\!& \left<0\right \vert {\mathfrak A}_{j_{1}}(\theta_{1})\; {\mathfrak B}(0)\;
{\mathfrak A}_{j_{M}}(\theta_{M})^{\dagger}{\mathfrak A}_{j_{M-1}}(\theta_{M-1})^{\dagger}
\cdots
{\mathfrak A}_{j_{2}}(\theta_{2})^{\dagger} \left\vert 0\right> \nonumber \\
&\!\!-\!\!& \left<0\right \vert {\mathfrak A}_{j_{1}}(\theta_{1})\; {\mathfrak B}(0)\left\vert 0\right>
\left< 0 \right\vert
{\mathfrak A}_{j_{M}}(\theta_{M})^{\dagger}{\mathfrak A}_{j_{M-1}}(\theta_{M-1})^{\dagger}
\cdots
{\mathfrak A}_{j_{2}}(\theta_{2})^{\dagger} \left\vert 0\right> \;,\nonumber
\eeq
which is ``connected" in the sense that the vacuum intermediate channel is subtracted 
\cite{BabKarow}. This Green's function can be thought of as $M-1$ incoming particles 
being absorbed by a vertex corresponding to the operator ${\mathfrak B}(0)$ which then
emits a single particle. Consider the pair of particles, with labels $1$ (the outgoing particle)
and $M$. Under crossing, these both become incoming particles, but with $\theta_{1}$
replaced by $\theta_{1}+\pi {\rm i}$. To see this, notice that this change preserves
all the relativistic invariants $(p_{j}\pm p_{j+1})^{2}$, $j=2,\dots,M-1$, but
it switches the two invariants $s_{1M}=(p_{1}+p_{M})^{2}$ and 
$t_{1M}=(p_{1}-p_{M})^{2}$. Thus
\beq
\left< \!\!\! \right. \!\!\!&0\!\!\!\!&\!\! \left. \right\vert {\mathfrak A}_{j_{1}}(\theta_{1})\; {\mathfrak B}(0)\;
{\mathfrak A}_{j_{M}}(\theta_{M})^{\dagger}{\mathfrak A}_{j_{M-1}}(\theta_{M-1})^{\dagger}
\cdots
{\mathfrak A}_{j_{2}}(\theta_{2})^{\dagger} \left\vert 0\right>_{\rm connected}  \nonumber \\ 
&\!\!=\!\!&
f^{\mathfrak B}(\theta_{1}+\pi{\rm i}, \theta_{1},\dots,\theta_{M})_{j_{1}\cdots j_{M}}\;.
\label{crossing1}
\eeq
Suppose that instead of switching the invariants $s_{1M}$ and $t_{1M}$, we switch
the invariants $s_{12}=(p_{1}+p_{2})^{2}$ and $t_{12}=(p_{1}-p_{2})^{2}$. Then we find
\beq
\left< \!\!\! \right. \!\!\!&0\!\!\!\!&\!\! \left. \right\vert {\mathfrak A}_{j_{1}}(\theta_{1})\; {\mathfrak B}(0)\;
{\mathfrak A}_{j_{M}}(\theta_{M})^{\dagger}{\mathfrak A}_{j_{M-1}}(\theta_{M-1})^{\dagger}
\cdots
{\mathfrak A}_{j_{2}}(\theta_{2})^{\dagger} \left\vert 0\right>_{\rm connected}  \nonumber \\ 
&\!\!=\!\!&
f^{\mathfrak B}(\theta_{2}, \theta_{1},\dots,\theta_{M}, \theta_{1}-\pi{\rm i})_{j_{2}\cdots j_{M} j_{1}}\;.
\label{crossing2}
\eeq
Comparing (\ref{crossing1}) and (\ref{crossing2}), the axiom follows.

\vspace{5pt}

\noindent
3. {\em Annihilation Pole Axiom}. This axiom is important
for relating form factors of $M$ particles to form factors
of $M-2$ particles. Though we will not apply $M>2$ form factors in this paper, we
mention the axiom 
anyway, since higher-point 
form factors should be eventually be of interest in our problem. The idea is the following:
let us again consider the form factor 
$f^{\mathfrak B}(\theta_{1},\dots,\theta_{M})_{j_{1}\cdots j_{M}}$. Now there is the possibility of
the $({M-1})^{\rm st}$ and $M^{\rm th}$ particles annihilating (they can be antiparticles)
at $\theta_{M}=\theta_{M-1}+\pi {\rm i}$ (so that $s_{M-1\,M}=(p_{M-1}+p_{M})^{2}$ 
will vanish). Thus
there must be a pole in the form factor which corresponds to this annihilation at 
$\theta_{M}=\theta_{M-1}+\pi {\rm i}$. There are two terms in the residue of this
pole: i) The $(M-1)^{\rm st}$ particle may scatter with
the $(M-2)^{\rm nd}$, $(M-3)^{\rm rd}$, $\dots$, $1^{\rm st}$ particles 
before annihilating the $M^{\rm th}$ particle,
or ii) the $(M-1)^{\rm st}$ particle may not scatter with any particles 
before annihilating the $M^{\rm th}$ particle. The axiom is
\beq
&\!\!{\rm i}\!\!&\!\!\left. {\rm Res} f^{\mathfrak B}(\theta_{1},\dots,\theta_{M})_{j_{1}\cdots j_{M}}\right\vert_{\theta_{M}=\theta_{M-1}+\pi{\rm i}}
=f^{\mathfrak B}(\theta_{1},\dots,\theta_{M-2})_{j_{1}\cdots j_{M-2}} \,C_{j_{M-1} j_{M}} \nonumber \\
&\!\!-\!\!&\!\!  
S^{k_{M-1} k_{1}}_{t_{1}j_{1}}(\theta_{1}-\theta_{M-1})
S^{t_{1} k_{2}}_{t_{2}j_{2}}(\theta_{2}-\theta_{M-1})\cdots
S^{t_{M-4} k_{M-3}}_{t_{M-3}j_{M-3}}(\theta_{M-3}-\theta_{M-1})   \nonumber
\\
&\!\! \times \!\! & \!\! S^{k_{M-3} k_{M-2}}_{j_{M-1}j_{M-2}}(\theta_{M-2}-\theta_{M-1})   
f^{\mathfrak B}(\theta_{1},\dots,\theta_{M-2})_{k_{1}\cdots k_{M-2}} \,C_{k_{M-1} j_{M}} \;, 
\label{axiom3}
\eeq
where $C$ is the charge-conjugation matrix. The normalization of the left-hand side
follows from the standard state normalization, {\em e.g.} $<\theta\vert\theta^{\prime}>
=2\pi\delta(\theta-\theta^{\prime})$.

\vspace{5pt}

\noindent
4. {\em Lorentz-Invariance Axiom}. If an operator ${\mathfrak B}(x)$ carries Lorentz spin s, the form
factors must
transform under a boost $\theta_{j}\rightarrow \theta_{j}+\alpha$ for all $j=1,\dots, M$ as
\beq
f^{\mathfrak B}(\theta_{1}+\alpha, \dots, \theta_{M}+\alpha)=e^{{\rm s}\alpha}
f^{\mathfrak B}(\theta_{1}, \dots, \theta_{M}) \;, \label{axiom4}
\eeq
We hope that the reader will have no trouble distinguishing spin s from center-of-mass energy
squared $s$.

\vspace{5pt}

\noindent
5. {\em Minimality Axiom}. Just as we assume
S-matrix elements have as much analyticity as possible, so we make a similar assumption
of form factors. In order to check the validity of this principle, all that can be done is
to compare with some perturbative method, which means either standard covariant
perturbation theory or
the $1/N$-expansion. In cases where we can find form factors, minimality stands up very well. If
a first guess for the form factor $f^{\mathfrak B}(\theta_{1}, \dots, \theta_{M})$ satisfies
the first four axioms, then so does
\beq
f_{\rm minimal}^{\mathfrak B}(\theta_{1}, \dots, \theta_{M})=
f^{\mathfrak B}(\theta_{1}, \dots, \theta_{M})
\frac{P_{M}(\{ \cosh(\theta_{j}-\theta_{k}) \})}{Q_{M}( \{  \cosh(\theta_{j}-\theta_{k}) \} )}\;,
\label{axiom5}
\eeq
where $P_{M}$ and $Q_{M}$ are symmetric polynomials. In this way, we can eliminate all the
poles in form factors, except those 
corresponding to bound states. In order for Axiom 3
to be satisfied:
\beq
\left.P_{M} \right\vert_{\theta_{M}=\theta_{M-1}+\pi{\rm i}}=P_{M-2}\;,\;\;
\left.Q_{M} \right\vert_{\theta_{M}=\theta_{M-1}+\pi{\rm i}}=Q_{M-2}\; . \nonumber
\eeq

The overall normalization of the form factors is not determined by these axioms The 
normalization can
be found for the case of current operators, which is what we shall
apply to the $(2+1)$-dimensional Yang-Mills theory.

\vspace{5pt}
\begin{center}
{\bf A3. Currents of the O($N$) nonlinear sigma model}
\end{center}
\vspace{5pt}

After this review of form-factor concepts, we next apply these ideas to the case of
two-particle form factors, which we shall find explicitly for currents of
the O$N$) sigma model. The two-particle form factor may be written as
\beq
f^{\mathfrak B}(\theta_{1},\theta_{2})_{j_{1} j_{2}}
=e^{-{\rm s}\theta_{1}} F^{\mathfrak B}(\theta_{2}-\theta_{1})_{j_{1}j_{2}} \;, \nonumber
\eeq
by Axiom 4. In terms of the function 
$F^{\mathfrak B}(\theta_{1},\theta_{2})_{j_{1}j_{2}}$, Axioms 1 and 2 are
\beq
F^{\mathfrak B}(\theta)_{k_{1}k_{2}}\!\!&\!\!=\!\!&\!\!e^{-{\rm s}\theta} S^{j_{1}j_{2}}_{k_{1}k_{2}}(\theta)
F^{\mathfrak B}(-\theta)_{j_{1}j_{2}}\;, \nonumber \\
F^{\mathfrak B}(2\pi{\rm i}-\theta)_{j_{2}j_{1}} 
\!\!&\!\!=\!\!&\!\!e^{{\rm s}\theta}F^{\mathfrak B}(\theta)_{j_{1}j_{2}}\;,
\label{useful-axioms}
\eeq
respectively.

Let us now recall the projectors $P_{0}$, $P_{S}$ and $P_{A}$ defined in 
(\ref{representation-projections}). We 
have already used the fact that these diagonalize the 
S-matrix acting on species tensors transforming as $N\otimes N$. We define
\beq
F^{\mathfrak B}_{0,S,A}(\theta)_{k_{1}k_{2}}=(P_{0,S,A})^{j_{1}j_{2}}_{k_{1}k_{2}}
F^{\mathfrak B}(\theta)_{j_{1}j_{2}} \;, \nonumber
\eeq
which allows us to rewrite (\ref{useful-axioms}) as 
\beq
F^{\mathfrak B}_{0,S,A}(\theta)
\!\!&\!\!=\!\!&\!\!e^{-{\rm s}\theta}S_{0,S,A}(\theta) F^{\mathfrak B}_{0,S,A}(-\theta) \;, \nonumber \\
F^{\mathfrak B}_{0,S}(\theta)=e^{{\rm s}\theta}F^{\mathfrak B}_{0,S}(2\pi{\rm i}-\theta)\!\!\!\!&,&
F^{\mathfrak B}_{A}(\theta)=-e^{{\rm s}\theta}F^{\mathfrak B}_{A}(2\pi{\rm i}-\theta)\;.
\label{useful-axioms1}
\eeq
We can solve the equations (\ref{useful-axioms1}) for the form factors, up to an overall
normalization by a contour-integration method \cite{KarowskiWeisz}. 

Consider a function $F(\theta)$ satisfying
\beq
F(2\pi{\rm i}-\theta)=-e^{{\rm s}\theta}F(\theta)\;,\;\; F(\theta)=e^{{\rm s}\theta} S_{A}(\theta)F(-\theta)\;.
\nonumber
\eeq 
Define the contour $C$ to be that from
$-\infty$ to $\infty$ and from $\infty+2\pi {\rm i}$ to $-\infty+2\pi{\rm i}$. Then
\beq
\ln F(\theta)\!\!&\!\!=\!\!&\!\!\int_{C} \frac{dz}{4\pi {\rm i}} \,\coth\frac{z-\theta}{2} \,\ln F(z) 
=\int_{-\infty}^{\infty} \frac{dz}{4\pi {\rm i}} \,\coth\frac{z-\theta}{2} \,
\ln \frac{F(z)}{F(z+2\pi{\rm i})} \nonumber 
\eeq
Differentiating this formula with respect to $\theta$ (we will explain why in a moment)
\beq
\frac{dF(\theta)}{d\theta}\!\!&\!\!=\!\!&\!\!\int_{-\infty}^{\infty}\frac{dz}{8\pi {\rm i}} \,
\frac{1}{\sinh^{2}\frac{z-\theta}{2}} 
\,\ln \frac{F(z)}{-e^{{\rm s}\theta}F(-z)}   \nonumber \\
\!\!&\!\!=\!\!&\!\!\int_{-\infty}^{\infty} \frac{dz}{8\pi {\rm i}} \,\frac{1}{\sinh^{2}\frac{z-\theta}{2}} 
\,\ln S_{A}(z) \;.\nonumber
\eeq
From our expression for $S_{A}$ in (\ref{the-S-matrix!}) and differentiating the integral formula
\beq
\int_{-\infty}^{\infty}\frac{dz}{4\pi {\rm i}}\, \coth\frac{z-\theta}{2}\, \sinh \frac{\xi z}{\pi{\rm i}}
=\frac{\sin^{2} \frac{\xi}{\pi}(\pi{\rm i}-\theta)}{\sinh \xi}-
\frac{1}{2\sinh \xi} \label{integral-formula}
\eeq
(which can be done using basic complex-integration methods) with respect to $\theta$, and
finally integrating $dF(\theta)/d\theta$ with respect to $\theta$, 
\beq
F(\theta)=G\exp \;2\int_{0}^{\infty} \frac{d\xi}{\xi}\, \frac{e^{\frac{-2\xi}{N-2}}-1}{e^{\xi}+1}\,
\frac{\sin^{2} \frac{\xi}{2\pi}(\pi{\rm i}-\theta)}{\sinh \xi} \;,\label{formula-for-F}
\eeq
where $G$ is a constant. The reason we had to differentiate and integrate with respect to
$\theta$ is that otherwise, the second term of (\ref{integral-formula}) will not lead to a convergent
answer.

Now we consider the form factors of the O($N$) current operator of the sigma model. This
operator is $J_{\mu}^{jk}=n^{j}\partial_{\mu}n^{k}-n^{k}\partial_{\mu}n^{j}$. By 
Hermiticity, translation invariance,
antisymmetry and Lorentz invariance
\beq
\left< 0 \right\vert J_{\mu}^{jk}(0) \left\vert \theta_{2},m,\theta_{1},l \right>
\!\!\!&\!\!=\!\!&\!\!{\rm i}G(\delta^{j}_{l}\delta^{k}_{m}-\delta^{j}_{m}\delta^{k}_{l})(p_{1}-p_{2})_{\mu}
\nonumber \\
\!\!&\!\! \times \!\!&\!\! \exp \;2\int_{0}^{\infty} \frac{d\xi}{\xi}\, \frac{e^{\frac{-2\xi}{N-2}}-1}{e^{\xi}+1}\,
\frac{\sin^{2} \frac{\xi}{2\pi}(\pi{\rm i}-\theta_{12})}{\sinh \xi} \;. \label{current-form-factor}
\eeq
This is the expression for the current-operator
form factor with as much analyticity as possible. The 
normalization of the right-hand side of (\ref{current-form-factor}) is obtained by the crossing 
relation (\ref{crossing1}) for the two-point case. We will fix this normalization for the
case of $N=4$, which is the case we wish to study further. It is not difficult, however, to
obtain the normalization for any $N$ using slightly different methods.

The normalization is fixed by examining the matrix elements of a charge operator. The eigenvalues
of this operator are fixed by symmetry considerations. In this way the value of the form
factor can be specified for particular rapidities. We will find that $G=1$.

As we already mentioned, the connection between the SU($2$)-valued
field $U$ and the unit four-vector $n$ is $U=n^{4}-{\rm i}n^{b}\sigma_{b}$. The relation between
the currents defined in (\ref{current-definition}) for the principal chiral sigma model and those
for the vector sigma model is therefore
\beq
j^{\rm L}_{\mu\;b}={\sqrt 2} \left( J^{4b}_{\mu}+\frac{1}{2}\epsilon_{bcd}J^{cd}_{\mu}  \right) \;,\;\;
j^{\rm R}_{\mu\;b}={\sqrt 2} \left( J^{4b}_{\mu}-\frac{1}{2}\epsilon_{bcd}J^{cd}_{\mu}  \right)\;.
\nonumber
\eeq
The left-handed charge density obeys the algebra
\beq
[j^{\rm L}_{0}(x^{1})_{b},j^{\rm L}_{0}(y^{1})_{c}]={\rm i}{\sqrt{2}}\epsilon^{bcd}\delta(x^{1}-y^{1})
j^{\rm L}_{\mu}(x^{1})_{d}\;, \nonumber
\eeq
and the left-handed charge is $Q_{b}=\int dx^{1}J^{L}_{0}(x^{1})_{b}$. The charge therefore
obeys the commutation relations
\beq
[Q^{\rm L}_{b},Q^{\rm L}_{c}]={\rm i}{\sqrt{2}}\epsilon^{bcd}
Q^{\rm L}_{d}\;, \nonumber
\eeq
Since the charge is in an orbital representation, the possible eigenvalues of $Q^{\rm L}_{3}$
are $0$ (isospin zero), $0,\pm {\sqrt{2}}$ (isospin one), etc. We find the same relations for the
right-handed charge.

From (\ref{form-factor-x}) and (\ref{current-form-factor}), we have
\beq
\left< 0 \right\vert \!\!\!&\!\!\!j\!\!\!&\!\!\!^{L,R}_{0}(x)_{b} \left\vert \theta_{2},j_{2},\theta_{1},j_{1}\right>=
{\rm i}{\sqrt 2}G\left( \delta_{j_{1} 4}\delta_{j_{2}b}  -
\delta_{j_{2} 4}\delta_{j_{1}b}  \pm \epsilon_{b j_{1} j_{2} }\right) 
m (\cosh \theta_{1} -\cosh \theta_{2}) \nonumber \\
\!\!&\!\!\times\!\!&\!\! 
\exp\{-{\rm i} m [ x^{0}(\cosh\theta_{1}+\cosh \theta_{2})-x^{1}(\sinh\theta_{1}+\sinh \theta_{2})
] \}\, F(\theta_{2}-\theta_{1}) \;, \label{useful-form-factor}
\eeq
where the plus or minus sign corresponds to the left-handed ($L$) or right-handed ($R$) current, respectively,
and
\beq
F(\theta)\!=\exp 2\int_{0}^{\infty} \frac{d\xi}{\xi}\, \frac{e^{-\xi}-1}{e^{\xi}+1}\,
\frac{\sin^{2} \frac{\xi(\pi{\rm i}-\theta)}{2\pi}}{\sinh \xi} 
\!=  \exp -\int_{0}^{\infty} \frac{d\xi}{\xi}\, \frac{e^{-\xi}}{\cosh^{2}\frac{\xi}{2}}\,
\sin^{2} \frac{\xi(\pi{\rm i}-\theta)}{2\pi}. \label{F-function}
\eeq

Now under crossing, the relation (\ref{crossing1}) for our simple two-point case yields
\beq
\left< \theta_{1},j_{1} \right\vert
 \!\!\!&\!\!\!j\!\!\!&\!\!\!^{L,R}_{0}(x)_{b} \left\vert \theta_{2},j_{2}
 \right>={\rm i}G{\sqrt{2}} \left( \delta_{j_{1} 4}\delta_{j_{2}b}  -
\delta_{j_{2} 4}\delta_{j_{1}b}  \pm \epsilon_{b j_{1} j_{2} }\right) 
m (\cosh \theta_{1} +\cosh \theta_{2}) \nonumber \\
\!\!&\!\!\times\!\!&\!\! 
\exp\{-{\rm i} m [ x^{0}(\cosh\theta_{1}-\cosh \theta_{2})-x^{1}(\sinh\theta_{1}-\sinh \theta_{2})
] \} \nonumber \\
\!\!&\!\! \times \!\! &\!\! F(\theta_{2}-\pi{\rm i}+\theta_{1}) \;, \label{crossed-form-factor}
\eeq
Integrating each side of (\ref{crossed-form-factor}) over $x^{1}$ yields
\beq
\left< \theta_{1},j_{1} \right\vert Q^{L,R}_{b} \left\vert \theta_{2},j_{2}
 \right>={\rm i}G{\sqrt{2}} \left( \delta_{j_{1} 4}\delta_{j_{2}b}  -
\delta_{j_{2} 4}\delta_{j_{1}b}  \pm \epsilon_{b j_{1} j_{2} }\right) \left< \theta_{1}\right\vert \left. \theta_{2} \right> \;. \nonumber
\eeq
This result yields the eigenvalues of $Q^{L,R}_{b}$ for isospin one, provided $G=1$.

Equation (\ref{current-form-factor}) with $G=1$
is a result of  Karowski and Weisz \cite{KarowskiWeisz}, who also checked its validity 
in the $1/N$-expansion. We shall use it to obtain an expression for the 
effective action of the electrostatic potential $\Phi$.

When an interaction is added to the action of an integrable model, the form factors will recieve
corrections \cite{Delfino}. The mass spectrum will be altered as well. We shall not work to high enough
order in $g_{0}^{\prime}$ for these effects to be taken into account. If higher-order calculations can be 
done, however, they will need to be included.


\begin{thebibliography}{xx}
\bibitem{PhysRevD71} P. Orland, Phys. Rev. {\bf D71} (2005) 054503. 
\bibitem{mandelstam} S. Mandelstam, Bull. Am. Phys. Soc. 
{\bf 22} (1977) 541; Phys. Rev. {\bf D19} (1979) 2391.
\bibitem{KarowskiWeisz} M. Karowski and P. Weisz, Nucl. Phys. {\bf B139} (1978) 455.
\bibitem{Griffin} P.A. Griffin, in {\bf Fourth International Workshop on Light-Cone Quantization
and Non-Perturbative Dynamics}, Polana Zgorzelisko, Poland 1994. Published in {Polana
Light-Cone Quantiz.} (1994) 240, {\bf hep-th/9410243}.
\bibitem{feynman} R.P. Feynman, Nucl. Phys. {\bf B188} (1981) 479.
\bibitem{orl-sem} P. Orland and G.W. Semenoff, Nucl. 
Phys. {\bf B576} (2000) 627.
\bibitem{orbit-space} P. Orland, {\bf hep-th/9607134} (1996).
\bibitem{kar-nair} D. Karabali and V.P. Nair, Nucl. Phys. {\bf B464} (1996) 135; Phys. Lett. {\bf B379} (1996) 141; 
D. Karabali, C. Kim and V.P. Nair 
{\bf B524} (1998) 661; Phys. Lett. {\bf B434} (1998) 103; Nucl. Phys. {\bf B566} (2000) 331, Phys. 
Rev. {\bf D64} (2001) 025011.
\bibitem{orl-gauge-inv} P. Orland, Phys. Rev. {\bf D70} (2004) 045014. 
\bibitem{Greensite} J.P. Greensite, Nucl. Phys. {\bf B166} (1980) 113.
\bibitem{leigh-min-yel} R.G. Leigh, D. Minic and A. Yelnikov, {\bf hep-th/0604060} (2006).
\bibitem{teper} M.J. Teper, Phys. 
Rev. {\bf D59} (1998) 014512.
\bibitem{CaselleGrinzaMagnoli} M. Caselle, P. Grinza and N. Magnoli, {\bf hep-th/0607014} (2006).
\bibitem{GreensiteReview} J. Greensite, Prog. Part. Nucl. Phys. {\bf 51} (2003) 1.
\bibitem{newpaper} P. Orland, {\bf hep-th/0608067} (2006), submitted to Physical Review D.
\bibitem{Zamolodchikov} A.B. Zamolodchikov and Al. B. Zamolodchikov, Nucl. Phys. {\bf B133}
(1978) 525. 
\bibitem{Creutz} M. Creutz, {\bf Quarks, Gluons 
and Lattices}, Cambridge
University Press (1983), Chapter 15.
\bibitem{FuNielsen} Y.K. Fu and H.B. Nielsen, Nucl. Phys. {B236} (1984) 167; Nucl. Phys. {B254} 
(1985) 127.
\bibitem{deconstruction} N. Arkani-Hamed, A.G. Cohen and H. Georgi, Phys. Rev. Lett. {\bf 86} (2001) 4557.
\bibitem{bal} A.P. Balachandran, G. Marmo, B.S. Skagerstam and A. Stern, {\bf Gauge Symmetries and Fibre Bundles: Applications to Particle Dynamics}, Lect. Notes Phys. {\bf 188} (1983), 1-140,
Springer-Verlag; A.M. Polyakov, Mod. Phys. Lett. {\bf A3} (1988) 325; H.B. Nielsen and D. Rohrlich,
Nucl. Phys. {\bf B299} (1988) 471; T. 
Jacobson, Phys. Lett. {\bf B216} (1989) 150; P. Orland, Int. J. Mod. Phys. {\bf A4} (1989) 3615.
\bibitem{Delfino} G. Delfino, G. Mussardo and P. Simonetti, Nucl. Phys. {\bf B473}(1996) 469. 
\bibitem {abda-wieg}  E. Abdalla, M.C.B. Abdalla and A. Lima-Santos,
Phys. Lett. {\bf B140} (1984) 71; P.B. Wiegmann; Phys. Lett. {\bf B142} (1984) 173.
\bibitem{pol-wieg} A.M. Polyakov and P.B.  Wiegmann, Phys. Lett. {\bf B131} (1983) 121;  P.B. 
Wiegmann, Phys. Lett. {\bf B141} (1984) 217.
\bibitem{Fateev-Kazakov-Wiegmann} V.A. Fateev, V.A. Kazakov and P.B. Wiegmann, Nucl. Phys.
{\bf B424} (1994) 505.
\bibitem{sigma} A.M. Polyakov, Phys. Lett. {\bf B59} (1975) 87; E. Brezin, J. Zinn-Justin and
J.C. Le Guillou, Phys, Rev. {\bf D14} (1976) 2615; E. Brezin, J. Zinn-Justin, Phys Rev. {\bf B14} (1976) 3110; W.A. Bardeen, B.W. Lee and R.E. Schrock, Phys. Rev. {\bf D14} (1976) 985; Y.Y.
Goldschmidt and E. Witten, Phys. Lett. {\bf 91B} (1980) 392.
\bibitem{pohlmeyer} K. Pohlmeyer, Commun. Math. Phys. {\bf 46} (1976) 207; M. L\"{u}scher and 
K. Pohlmeyer, Nucl. Phys. {\bf B137} (1978) 46.
\bibitem{arafyeva} I.Ya Arafeva, P.P. Kulish, E.R. Nisimov and S.J. 
Pacheva, preprint LOMI-E-1-1978 (1977); A.M. Polyakov, Phys. Lett. {\bf B72} (1977) 224.
\bibitem{stanley} H.E. Stanley, Phys. Rev. {\bf 176} (1968) 718.
\bibitem{McCoy+co} B.M. McCoy, C.A. Tracy and T.T. Wu, Phys. Rev. Lett. {\bf 38} (1977) 793;
M. Sato, T. Miwa and M. Jimbo, report Kyoto RIMS-225(1977); B. Berg, M. Karowski and P. Weisz, Phys. Rev.
{\bf D19} (1979) 2477.
\bibitem{zam+zam}A.B. Zamolodchikov 
and Al.B. Zamolodchikov, Ann. Phys. {\bf 120} (1979) 253.
\bibitem{KTTW} M. Karowski, H.J. Thun, T.T. Truong and P. Weisz, Phys. Lett. {\bf B67} (1977) 321.
\bibitem{zamS-G} A.B. Zamolodchikov, Commun. Math. Phys.  {\bf 55} (1977) 183.
\bibitem{KarowskiThun} M. Karowski and H.J. Thun, Nucl. Phys. {\bf B130} (1977) 295. 
\bibitem{BergReview} B. Berg, in {\bf International Colloquium of Complex Analysis, Microlocal
Calculus and Relativistic Quantum Field Theory}, Sept. 3-13, 1979 Les Houches, France (1980).
\bibitem{W+W} E.T. Whittaker and G.N. Watson, {\bf A Course of Modern Analysis}, Cambridge University Press, Chapter 12, pg. 249.
\bibitem{Weisz} P. Weisz, Phys. Lett. {\bf B67} (1977) 179.
\bibitem{1/N} B. Berg, M. Karowski, V. Kurak and P. Weisz, Phys. Lett. {\bf B76} (1978) 502.
\bibitem{CDD} L. Castillejo, R.H. Dalitz and F.J. Dyson, Phys. Rev. {\bf 101} (1956) 453.
\bibitem{ShankarWitten} R. Shankar and E. Witten, Phys. Rev. {\bf D17} (1978) 2134.
\bibitem{VergelesGryanik} S. Vergeles and V. Gryanik, Sov. Journ. Nucl. Phys. {\bf 23} (1976) 704.
\bibitem{Smirnov} F.A. Smirnov, preprint LOMI-E-3-1986 (1986); J. Phys. {\bf A19} (1986) L575; 
A.N. Kirillov and F. A. Smirnov, Phys. Lett. {\bf B198} (1987) 506; F.A. Smirnov, Int. J. Mod. Phys. {\bf A9} (1994) 5121.
\bibitem{SmirnovBook} F.A. Smirnov, {\bf Form Factors in Completely Integrable Models of Quantum Field Theory}, Adv. Series in Math. Phys. {\bf 14}, World Scientific (1992).
\bibitem{BabKarow} H. Babujian, A. Fring, M. Karowski and A. Zapletal, Nucl. Phys. {\bf B538} (1999) 535; H. Babujian and M. Karowski, Nucl. Phys. {\bf B620} (2002) 407.
\bibitem{FonsecaZamolodchikov} P. Fonseca and A.B. Zamolodchikov, J. Stat. Phys, {\bf 110} (2003) 527.
\bibitem{BhaseenTsvelick} M.J. Bhaseen and A.M. Tsvelik, in {\bf From Fields to Strings; Circumnavigating Theoretical Physics}, Ian Kogan
memorial volumes, Vol. 1 (2004), pg. 661,
M. Shifman, A. Vainshtein and J. Wheater ed., {\bf cond-mat/0409602}
\bibitem{EsslerKonik} F.H.L. Essler and R. Konik, ibid., pg. 684, {\bf cond-mat/0412421}.
\bibitem{KWatson} K.M. Watson, Phys. Rev., {\bf 95} (1954) 228.
\end{thebibliography}
\end{document}